\documentclass[prb,twocolumn,showpacs,preprintnumbers,superscriptaddress,longbibliography,dvipsnames]{revtex4-2}

\usepackage[utf8]{inputenc}
\usepackage{amsmath}
\usepackage{amssymb}
\usepackage{mathtools}
\usepackage{graphicx}
\usepackage{bm}
\usepackage{tikz}
\usetikzlibrary{shapes.geometric, arrows.meta}
\usepackage{booktabs}
\usepackage{hyperref}
\usepackage{cleveref}
\usepackage{multirow}
\usepackage{changepage}

\definecolor{dark-red}{rgb}{0.4,0.15,0.15}
\definecolor{dark-blue}{rgb}{0.15,0.15,0.4}
\definecolor{medium-blue}{rgb}{0,0,0.5}
\hypersetup{
	colorlinks,
	linkcolor={dark-blue},
	citecolor={dark-blue},
	urlcolor={medium-blue}
}

\newcommand{\Nf}{N_\mathrm{f}}
\renewcommand{\i}{\text{i}}

\begin{document}

\title{Continuous symmetry analysis and systematic identification of candidate order parameters for interacting fermion models}

\author{Cheng-Hao He}
\affiliation{Key Laboratory of Artificial Structures and Quantum Control (Ministry of Education), School of Physics and Astronomy, Shanghai Jiao Tong University, Shanghai 200240, China}
\author{Yi-Zhuang You}
\email{yzyou@physics.ucsd.edu}
\affiliation{Department of Physics, University of California, San Diego, CA 92093, USA}
\author{Xiao Yan Xu}
\email{xiaoyanxu@sjtu.edu.cn}
\affiliation{Key Laboratory of Artificial Structures and Quantum Control (Ministry of Education), School of Physics and Astronomy, Shanghai Jiao Tong University, Shanghai 200240, China}
\affiliation{Hefei National Laboratory, Hefei 230088, China}
\date{\today}

\begin{abstract}
Symmetry plays a central role in modern physics, from classifying quantum states to characterizing phases of matter through spontaneous symmetry breaking. In interacting fermionic systems with multiple internal degrees of freedom, however, determining the full continuous symmetry group and classifying possible order parameters remain challenging. In this work, we present a systematic framework for analyzing continuous symmetries and identifying candidate order parameters in such systems. By mapping the Hamiltonian to a Majorana representation, we obtain the generators of continuous symmetries from the Lie algebra of operators that commute with the Hamiltonian. We then identify the structure of this Lie algebra using the theory of semisimple Lie algebras. Building on representation theory, we further develop a systematic method for exhaustively enumerating candidate order parameters. By decomposing the exterior-power representations induced by the symmetry algebra on the Majorana space and incorporating discrete lattice symmetries, we classify these order parameters according to the symmetries they break. To demonstrate the power of the framework, we first apply it to the single layer Hubbard model on a honeycomb lattice as a benchmark and recover the well-known $\mathrm{SO}(4)$ symmetry with $\mathfrak{su}(2)\oplus\mathfrak{su}(2)$ symmetry algebra. We then apply it to a bilayer spin-$1/2$ fermion model on a honeycomb lattice with Heisenberg exchange and density-density interlayer couplings, uncovering a $\mathrm{Spin}(5) \times \mathrm{U}(1)/\mathbb{Z}_2$ symmetry with $\mathfrak{so}(5)\oplus\mathfrak{u}(1)$ symmetry algebra. Within this setting, we systematically classify all candidate bilinear order parameters and reveal a rich landscape of potentially competing phases. The same framework has also been applied to a closely related bilayer model with pure Heisenberg interlayer coupling, which possesses an $\mathrm{SU}(2) \times \mathrm{SU}(2) \times \mathrm{SU}(2)/\mathbb{Z}_2$ symmetry with $\mathfrak{su}(2)\oplus\mathfrak{su}(2)\oplus\mathfrak{su}(2)$ symmetry algebra~\cite{He2026a}.
\end{abstract}

\maketitle

\section{Introduction}
\label{sec:intro}

Symmetry stands as a cornerstone in our understanding of the physical universe. It not only provides the mathematical language to describe the underlying structure of nature, but also serves as a guiding principle for discovering fundamental physical laws. From the conservation laws in classical mechanics established via Noether's theorem to the gauge symmetries underpinning the Standard Model of particle physics, symmetry considerations are ubiquitous. In condensed matter physics, symmetries constrain electronic band structures, dictate selection rules, and provide the foundation for classifying topological phases~\cite{dresselhaus_group_2010}.

A central paradigm in the study of quantum and classical phases of matter is spontaneous symmetry breaking (SSB). The Landau theory of phase transitions describes the transition from a high-symmetry disordered phase to a low-symmetry ordered phase in terms of a local order parameter. Such an order parameter typically transforms under a non-trivial irreducible representation (irrep) of the underlying symmetry group of the Hamiltonian. Consequently, identifying the correct order parameter is crucial for characterizing the ordered phase and understanding the microscopic mechanism driving the phase transition.

For simple models, the symmetry group and potential order parameters can often be deduced largely by inspection. However, for complex many-body systems endowed with multiple internal degrees of freedom---such as spin, orbital, valley, and layer indices---analyzing the intricate symmetry structure and systematically enumerating all allowed order parameters becomes a formidable challenge. The complex interplay between these diverse degrees of freedom can give rise to enlarged or hidden emergent symmetries, which are often not immediately apparent in the standard complex fermion basis~\cite{You2015,bultinck_ground_2020,bernevig_twisted_2021,zhang_unified_1997,wu_exact_2003,li_signproblemfree_2019}. Related symmetry-based approaches have been recently employed to classify mass terms and order parameters in continuum Dirac systems and lattice models~\cite{Herbut2023,Han2024,rein2025phase}, as well as to constrain phase diagrams using antiunitary symmetries~\cite{rein2025phase}.

In this paper, we propose a systematic and algorithmic framework to address this challenge for Hamiltonians with local operators. We start by expressing the many-fermion Hamiltonian in terms of Majorana fermions, wherein continuous symmetries naturally manifest as orthogonal transformations. The problem of identifying the continuous symmetry group is thereby reduced to finding the corresponding Lie algebra, which consists of the antisymmetric matrices that commute with the Hamiltonian tensor. We then employ standard mathematical tools from the theory of semisimple Lie algebras---specifically Cartan subalgebras, root systems, and Dynkin diagrams---to identify the Lie algebra and its associated Lie group.

Once the symmetry group is established, we proceed to identify candidate order parameters. This is achieved by constructing the induced exterior-power representations of the Lie algebra on the spaces of fermion bilinears (or higher-order operators) and subsequently decomposing them into irreducible representations. To this end, we introduce a computational approach utilizing intertwiners to perform the representation decomposition of the Lie algebra, thereby obtaining the invariant subspaces of the connected component of the Lie group. Finally, acting on these invariant subspaces, we incorporate discrete symmetries to obtain the candidate order parameters.

To illustrate the utility of our proposed method, we first apply it to the Hubbard model on a honeycomb lattice as a benchmark, recovering the well-known $\mathrm{SO}(4)$ symmetry with $\mathfrak{su}(2)\oplus\mathfrak{su}(2)$ symmetry algebra. We then apply it to a bilayer spin-$1/2$ fermion model defined on a honeycomb lattice with both Heisenberg exchange and density-density interlayer couplings. We demonstrate that the model possesses an $\mathrm{Spin}(5) \times \mathrm{U}(1)/\mathbb{Z}_2$ symmetry with $\mathfrak{so}(5)\oplus\mathfrak{u}(1)$ symmetry algebra. Furthermore, we systematically classify all possible bilinear order parameters associated with the internal degrees of freedom within a unit cell for this model. This classification reveals a rich landscape of potentially competing quantum phases. The same framework also applies to a closely related model with a pure Heisenberg interlayer coupling, which possesses an $\mathrm{SU}(2) \times \mathrm{SU}(2) \times \mathrm{SU}(2)/\mathbb{Z}_2$ (abbreviated as $\mathrm{SU}(2)^3/\mathbb{Z}_2$ in the following) symmetry with $\mathfrak{su}(2)\oplus\mathfrak{su}(2)\oplus\mathfrak{su}(2)$ (abbreviated as $3\,\mathfrak{su}(2)$ in the following) symmetry algebra; the full symmetry analysis and order parameter classification for that model are presented in a companion paper~\cite{He2026a}. The systematic classification of all candidate order parameters is essential for establishing symmetric mass generation (SMG)~\cite{wang2022symmetric,You2018,tong2022comments,Zeng2022,Lu2023,xu2021green,You2018from}, as it requires demonstrating the absence of long-range order for every symmetry-breaking channel in the gapped phase. In the companion paper~\cite{He2026a}, quantum Monte Carlo simulations show that the $\mathrm{SU}(2)^3/\mathbb{Z}_2$ model exhibits a direct transition from the Dirac semimetal to an SMG phase, while the $\mathrm{Spin}(5) \times \mathrm{U}(1)/\mathbb{Z}_2$ model features an intermediate excitonic phase between the Dirac semimetal and the SMG phase.

The remainder of this paper is organized as follows. In Sec.~\ref{sec:symmetry_analysis}, we detail the general framework for continuous symmetry analysis using the Majorana representation. In Sec.~\ref{sec:application_symmetry}, we apply this framework to the Hubbard model and the bilayer spin-$1/2$ model with Heisenberg exchange and density-density interlayer couplings to explicitly determine their symmetry groups. Sec.~\ref{sec:order_parameter_search} describes the algorithmic method for identifying candidate order parameters via irreducible decomposition. In Sec.~\ref{sec:application_op}, we present the complete classification of candidate order parameters for the Hubbard model and the bilayer model. Finally, we summarize our findings in Sec.~\ref{sec:conclusion}.

\section{Continuous Symmetry Analysis}
\label{sec:symmetry_analysis}

The goal of this section is to recast the search for continuous symmetries of a fermionic Hamiltonian $H$ as an algebraic problem. In the Majorana representation, continuous symmetry operations act as real orthogonal transformations on the Majorana operators, and their infinitesimal generators are represented by antisymmetric matrices. The problem therefore reduces to finding the generators that commute with $H$. Once these generators are known, the corresponding Lie algebra and hence the continuous symmetry group can be identified.

\subsection{Majorana Representation}

In this work, we focus on interacting fermion models whose Hamiltonians can be decomposed into local terms:
\begin{equation}
    H=\sum_{b}J_{b}H_{b},
    \label{eq:Hamiltonian}
\end{equation}
where $H_{b}$ represents a local operator and $J_{b}$ is the corresponding coupling strength. As a simple example, consider a lattice model with nearest-neighbor hopping and onsite interactions:
\begin{equation}
    H=-\sum_{\left\langle i,j\right\rangle }t_{i,j}c_{i}^{\dagger}c_{j}+\sum_{i}U_{i}H_{I,i},
    \label{eq:Hamiltonian-example}
\end{equation}
where $i$ labels lattice sites, $\left\langle i,j\right\rangle$ denotes nearest-neighbor pairs, $t_{i,j}$ is the hopping amplitude, $U_i$ is the interaction strength, and $H_{I,i}$ is the local interaction operator. We may then define a local Hamiltonian $H_{b}$ for each bond $b = \langle i,j \rangle$ that collects all terms acting on that bond:
\begin{equation}
    H_{b}=t_{i,j}c_{i}^{\dagger}c_{j}+\frac{U_{i}}{z}H_{I,i}+\frac{U_{j}}{z}H_{I,j},
    \label{eq:Hamiltonian-local}
\end{equation}
where $z$ is the coordination number of the lattice. This construction ensures that $H = \sum_{b} H_{b}$ reproduces the full Hamiltonian.

To analyze symmetries systematically and place the particle-hole degree of freedom on the same footing as the other internal degrees of freedom, we work in the Majorana representation. Consider a subsystem with $\Nf$ local complex fermion modes $c_m$ ($m=1,\dots,\Nf$) appearing in the local Hamiltonian $H_b$, where the index $m$ includes all relevant spatial and internal labels, such as spin, layer, and sublattice. We introduce $2\Nf$ Majorana fermion operators defined as
\begin{equation}
    \begin{aligned}
    \gamma_{1,m} &= c_{m}^{\dagger} + c_{m}, \\
    \gamma_{2,m} &= \i(c_{m}^{\dagger} - c_{m}),
    \end{aligned}
    \label{eq:majorana_def}
\end{equation}
which can be inverted as
\begin{equation}
    \begin{aligned}
    c_{m}^{\dagger} &= \frac{1}{2}(\gamma_{1,m} - \i\gamma_{2,m}), \\
    c_{m} &= \frac{1}{2}(\gamma_{1,m} + \i\gamma_{2,m}).
    \end{aligned}
\end{equation}
These Majorana operators are Hermitian ($\gamma_{p,m} = \gamma_{p,m}^\dagger$) and satisfy the Clifford algebra anticommutation relations:
\begin{equation}
    \{\gamma_{p,m}, \gamma_{q,m^{\prime}}\} = 2\delta_{p q}\delta_{m m^{\prime}},
    \label{eq:majorana_anticommutation}
\end{equation}
where $p,q \in \{1,2 \}$. By flattening the pair of indices $(p,m)$, we assemble these operators into a single $2\Nf$-component vector $\bm{\gamma} = (\gamma_1, \dots, \gamma_{2\Nf})^T$, whose composite index runs from $1$ to $2\Nf$.

In this basis, a generic fermionic Hamiltonian becomes a polynomial in the Majorana operators. For instance, a local Hamiltonian composed of two-body ($H_{0,b}$) and four-body ($H_{I,b}$) terms takes the form:
\begin{gather}
    H_{0,b}=\sum_{\alpha,\beta}H_{\alpha\beta}^{\left(0,b\right)}\gamma_{\alpha}\gamma_{\beta},\\H_{I,b}=\sum_{\alpha,\beta,\mu,\nu}H_{\alpha\beta\mu\nu}^{\left(I,b\right)}\gamma_{\alpha}\gamma_{\beta}\gamma_{\mu}\gamma_{\nu},
\end{gather}
where the composite indices $\alpha,\beta,\mu,\nu \in \{1,\cdots, 2\Nf\}$, and the coefficient tensors $H^{(0,b)}$ and $H^{(I,b)}$ may be taken to be fully antisymmetric without loss of generality because the Majorana operators anticommute.

\subsection{Lie Algebra of Continuous Symmetries}
\label{sec:lie_algebra_method}

A continuous linear transformation of the Majorana fermions, given by $\tilde{\gamma}_\nu = \sum_\mu \gamma_\mu A_{\mu\nu}$, preserves the canonical anticommutation relations~\eqref{eq:majorana_anticommutation} if and only if the transformation matrix $A$ is orthogonal, namely $A A^T = I$. The continuous symmetry group of an $\Nf$-mode fermionic system is therefore a subgroup of $\mathrm{O}(2\Nf)$, and its generators form a subalgebra of $\mathfrak{so}(2\Nf)$.

Let $G \subseteq \mathrm{O}(2\Nf)$ denote the continuous symmetry group of the Hamiltonian $H$, and let $\mathfrak{g} \subseteq \mathfrak{so}(2\Nf)$ be its associated Lie algebra. Any element $X \in \mathfrak{g}$ generates a continuous transformation $A = e^{\theta X} \in G$. Requiring the Hamiltonian to be invariant under this transformation, $A H A^T = H$, leads at infinitesimal order to the commutation relation $[H, X] = 0$.

More precisely, the infinitesimal action of a generator $X$ on a single Majorana operator is given by $\delta \gamma_\nu = \sum_\mu \gamma_\mu X_{\mu\nu}$. Its action on a product of Majorana operators is the induced action on the corresponding exterior power. For a generic $m$-body term in the Hamiltonian, $H^{(m)} = \sum_{\{\alpha\}} H_{\alpha_1 \dots \alpha_m} \gamma_{\alpha_1} \dots \gamma_{\alpha_m}$, the commutation condition $[H^{(m)}, X]=0$ is equivalent to the tensor equation
\begin{equation}
     \sum_{k=1}^m \sum_{\alpha_k'} H_{\alpha_1 \dots \alpha_k' \dots \alpha_m} X_{\alpha_k'\alpha_k} = 0.
    \label{eq:commutation-condition}
\end{equation}
For this identity to hold, the fully antisymmetrized coefficients of the resulting Majorana polynomial must vanish. In practice, we expand $X$ in a complete basis of $\mathfrak{so}(2\Nf)$, denoted by $\{x_\mu\}$, such that $X = \sum_\mu a_\mu x_\mu$. Substituting this expansion into the commutation condition yields a homogeneous linear system for the real coefficients $a_\mu$. The null space of this system is precisely the Lie algebra $\mathfrak{g}$ commuting with the Hamiltonian.

Before summarizing the procedure, it is crucial to distinguish between the symmetry algebra of a local Hamiltonian $H_b$ and the global symmetry algebra of the full Hamiltonian $H = \sum_b H_b$. Any generator that commutes with every local term $H_b$ also commutes with $H$. The converse, however, need not hold automatically when one attempts to extend local generators to the entire lattice: geometrical frustration can obstruct such an extension. For the local bond Hamiltonians considered in Sec.~\ref{sec:application_symmetry}, the lattice is bipartite, so every bond connects two different sublattices and the local symmetries can be extended consistently to the full system. In that case, the continuous symmetry algebra of $H_b$ coincides with that of $H$. On a frustrated lattice, such as the triangular lattice, neighboring bonds can impose incompatible constraints, and the global symmetry algebra may then be a proper subalgebra of the local one.

\begin{figure}[!htp]
    \centering
    \begin{tikzpicture}[
  node distance=1.2cm,
  every node/.style={font=\footnotesize},
  startstop/.style={
    rectangle,
    rounded corners,
    minimum width=4em,
    text width=5cm,
    text centered,
    inner sep=1.2ex,
    draw
  },
  io/.style={
    trapezium,
    trapezium left angle=75,
    trapezium right angle=105,
    minimum width=4em,
    text width=5cm,
    text centered,
    inner sep=1.2ex,
    draw
  },
  process/.style={
    rectangle,
    minimum width=4em,
    text width=5.5cm,
    text centered,
    inner sep=1.2ex,
    draw
  },
  decision/.style={
    diamond,
    aspect=2,
    text centered,
    inner sep=0.5ex,
    draw
  },
  arrow/.style={-{LaTeX}, thick}
]

  \node[startstop] (start) {Hamiltonian $H$};
  \node[process, below of=start] (maj) {Determine local degrees of freedom and express $H$ in Majorana representation with basis dimension $2N$};
  \node[process, below of=maj,yshift=-0.15cm] (basis) {Construct complete basis for $\mathfrak{so}(2N)$};
  \node[process, below of=basis] (solve) {Solve $[H, \mathfrak{g}] = 0$ to identify Lie subalgebra $\mathfrak{g}$};
  \node[process, below of=solve, yshift=-0.0cm] (decomp) {Decompose $\mathfrak{g}$ into a direct sum of ideals};
  
  \node[decision, below of=decomp, yshift=-0.15cm] (semi) {Semisimple?};
  
  \node[process, right of=semi, xshift=2.0cm, text width=2.5cm] (abelian) {Identify as Abelian factor (e.g., $\mathfrak{u}(1)$)};
  \node[process, below of=semi, yshift=-0.4cm] (cartan) {Identify Cartan subalgebra $\mathfrak{h} \subset \mathfrak{g}$};
  \node[process, below of=cartan] (roots) {Diagonalize adjoint representations to extract roots};
  \node[process, below of=roots] (dynkin) {Construct root system \& Dynkin diagram};
  
  \node[process, below of=dynkin, yshift=0.1cm] (classify) {Classify the Lie algebra $\mathfrak{g}$};
  \node[startstop, below of=classify] (end) {Classify continuous symmetry group $G$};
  
  \draw[arrow] (start) -- (maj);
  \draw[arrow] (maj) -- (basis);
  \draw[arrow] (basis) -- (solve);
  \draw[arrow] (solve) -- (decomp);
  \draw[arrow] (decomp) -- (semi);
  
  \draw[arrow] (semi) -- node[above] {No} (abelian);
  \draw[arrow] (semi) -- node[right] {Yes} (cartan);
  
  \draw[arrow] (cartan) -- (roots);
  \draw[arrow] (roots) -- (dynkin);
  
  \draw[arrow] (dynkin) -- (classify);
  \draw[arrow] (abelian) |- (classify);
  \draw[arrow] (classify) -- (end);
\end{tikzpicture}
    \caption{Flowchart summarizing the continuous symmetry analysis of a many-body Hamiltonian.}
    \label{fig:symmetry-analysis}
\end{figure}
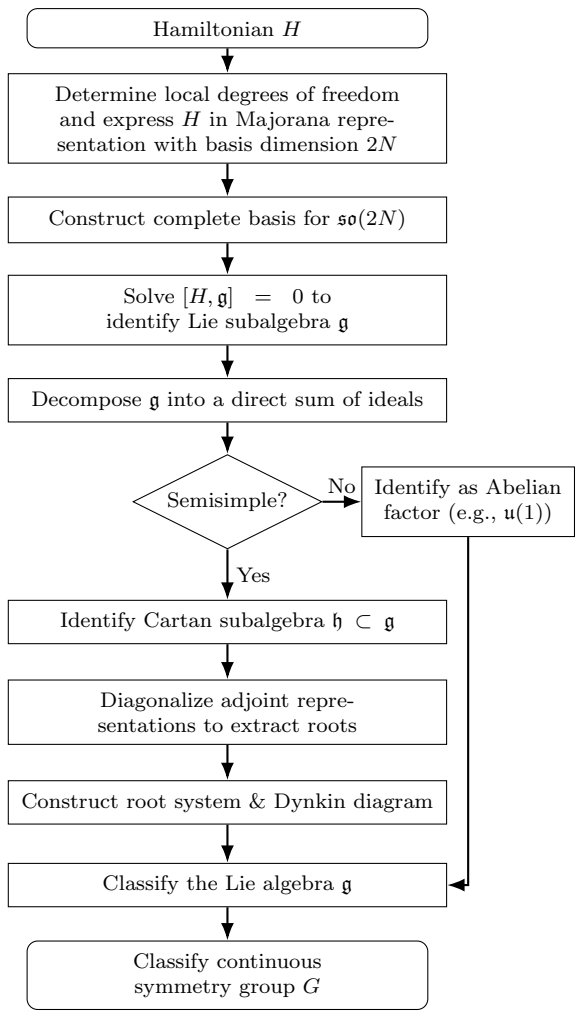

The complete procedure for analyzing continuous symmetries is summarized as follows (see also the flowchart in Fig.~\ref{fig:symmetry-analysis}):
\begin{enumerate}
    \item \textbf{Setup}: Determine the dimension $2\Nf$ of the local Majorana basis and express the local term $H_b$ of the Hamiltonian as a polynomial tensor in the Majorana operators.
    \item \textbf{Solve for Lie Algebra}: Construct a complete basis for $\mathfrak{so}(2\Nf)$. Identify the Lie subalgebra $\mathfrak{g} \subseteq \mathfrak{so}(2\Nf)$ that commutes with $H_b$ by solving the system of linear equations derived from the commutation conditions. Determine the global symmetry algebra of the total Hamiltonian $H$ by analyzing lattice frustration.
    \item \textbf{Identify Structure}: Analyze the algebraic structure of $\mathfrak{g}$ to formally classify the continuous symmetry group. In typical physical systems, this Lie algebra decomposes into a direct sum of semisimple and abelian components. 
    \begin{itemize}
        \item Verify whether the resulting Lie algebra $\mathfrak{g}$ (or its components) is semisimple.
        \item Identify a Cartan subalgebra $\mathfrak{h} \subset \mathfrak{g}$, which acts as the maximal commuting subalgebra.
        \item Diagonalize the adjoint representations of the elements in $\mathfrak{h}$ to extract the associated roots.
        \item Construct the root system and deduce its corresponding Dynkin diagram.
    \end{itemize}
    \item \textbf{Result}: The Dynkin diagram uniquely classifies the semisimple sector of the Lie algebra into standard families (such as $A_n, B_n, C_n, D_n$). Combining this identification with any independent abelian factors (e.g., $\mathfrak{u}(1)$) fully delineates the Lie algebra of the continuous symmetry group.
\end{enumerate}

This framework systematically exposes the symmetries of a model without relying on empirical assumptions or intuition, offering a robust technique for uncovering hidden symmetries in intricately coupled quantum systems. Strictly speaking, the Lie algebra approach inherently limits our analysis to the connected component of the symmetry group containing the identity element. Furthermore, distinct connected Lie groups can share the same Lie algebra, as is the case for $\mathrm{SU}(2)$ and $\mathrm{SO}(3)$, or more generally $\mathrm{SU}(2)\times\mathrm{SU}(2)$ and $\mathrm{SO}(4)\cong\mathrm{SU}(2)\times\mathrm{SU}(2)/\mathbb{Z}_2$. Given a Lie algebra $\mathfrak{g}=\bigoplus_i \mathfrak{g}_i$, the faithful symmetry group acting on the Hilbert space takes the form $G=\tilde{G}/\Gamma$, where $\tilde{G}=\prod_i \tilde{G}_i$ is the simply connected (universal covering) group and $\Gamma\subseteq Z(\tilde{G})$ is the subgroup of the center acting trivially on the physical Hilbert space. Determining $\Gamma$ requires examining which center elements act trivially, a calculation that depends on the specific representation content of the Hilbert space (see, e.g., Sec.~\ref{sec:application_symmetry} and the companion paper~\cite{He2026a}). Nevertheless, for many physical purposes, such as classifying the representations of local candidate order parameters, knowledge of the Lie algebra is practically sufficient.

\subsection{Application} \label{sec:application_symmetry}

To demonstrate our method, we apply it to two specific interacting fermion systems on the honeycomb lattice: the Hubbard model and a bilayer spin-$1/2$ model with Heisenberg exchange and density-density interlayer couplings.

\subsubsection{Hubbard Model}

The Hamiltonian of the Hubbard model is given by
\begin{gather} \label{eq:Hubbard-model}
    H=H_{0}+H_{I},\\H_{0}=-t\sum_{\left\langle i,j\right\rangle ,\sigma}c_{i,\sigma}^{\dagger}c_{j,\sigma}+\mathrm{h.c.,}\\H_{I}=U\sum_{i}n_{i\uparrow}n_{i\downarrow},
\end{gather}
where $U$ is the on-site interaction strength, and $n_{i\sigma}=c_{i\sigma}^\dagger c_{i\sigma}$ is the number operator for spin $\sigma\in\{\uparrow,\downarrow\}$ at site $i$. The associated local Hamiltonian $H_b$ defined on a bond $b=\langle i, j \rangle$ encapsulates both the hopping and the interaction terms:
\begin{equation} \label{eq:Hubbard-local}
    H_{b}=-t\left(c_{i}^{\dagger}c_{j}+c_{j}^{\dagger}c_{i}\right)+\frac{U}{z}\left(n_{i\uparrow}n_{i\downarrow}+n_{j\uparrow}n_{j\downarrow}\right).
\end{equation}
Thus, the local degrees of freedom consist of the two sites on a bond together with their particle-hole and spin indices
\begin{equation} \label{eq:DOF-bond}
    \underset{\text{bond}}{\begin{bmatrix}A\\
B
\end{bmatrix}}\otimes\underset{\text{particle-hole}}{\begin{bmatrix}c^{\dagger}\\
c
\end{bmatrix}}\otimes\underset{\text{spin}}{\begin{bmatrix}\uparrow\\
\downarrow
\end{bmatrix}}.
\end{equation}
This results in a Majorana space of dimension $2\Nf=2^3=8$.

To perform the symmetry analysis, we construct a basis for generators acting on the Majorana space. Since each degree of freedom in Eq.~\eqref{eq:DOF-bond} is two-dimensional, it is natural to use tensor products of Pauli matrices. Let $\sigma(0) = \mathrm{I}_{2\times2}$ be the identity matrix, and let $\sigma(1)=\sigma_x$, $\sigma(2)=\sigma_y$, and $\sigma(3)=\sigma_z$ be the standard Pauli matrices. We define the rank-3 tensor product
\begin{equation}
    \sigma(i,j,k) = \sigma(i) \otimes \sigma(j) \otimes \sigma(k),
\end{equation}
where the tensor-product order matches the degrees of freedom in Eq.~\eqref{eq:DOF-bond}: (bond) $\otimes$ (particle-hole) $\otimes$ (spin).

The corresponding Lie algebra $\mathfrak{so}(8)$ consists of all $8\times 8$ antisymmetric matrices. In this Pauli basis, antisymmetry requires an odd number of $\sigma(2)$ factors, since $\sigma(2)$ is the only antisymmetric Pauli matrix. The basis of $\mathfrak{so}(8)$ therefore contains $28$ matrices $\sigma(i,j,k)$ with either one or three indices equal to $2$.

Expressing the local Hamiltonian in Eq.~\eqref{eq:Hubbard-local} in the Majorana basis and imposing the commutation condition in Eq.~\eqref{eq:commutation-condition}, we obtain a $6$-dimensional Lie algebra $\mathfrak{g}$ with basis
\begin{equation} \label{eq:Hubbard-lie-algebra-basis}
    \begin{aligned}
        \begin{cases}
            x_{1}=\sigma\left(0,0,2\right)\\
            x_{2}=\sigma\left(0,2,3\right)\\
            x_{3}=\sigma\left(0,2,1\right)
            \end{cases},\begin{cases}
            x_{4}=\sigma\left(0,2,0\right)\\
            x_{5}=\sigma\left(3,3,2\right)\\
            x_{6}=\sigma\left(3,1,2\right)
            \end{cases}.
    \end{aligned}
\end{equation}
These basis matrices are orthonormal under the normalized inner product $\langle x_i,x_j \rangle = \frac{1}{8}\mathrm{Tr}(x_i^{\dagger} x_j)$. Evaluating the commutators $[x_i, x_j]$ shows that the algebra decomposes into a direct sum of two ideals,
\begin{equation}
    \mathfrak{g} = \mathfrak{g}_1 \oplus \mathfrak{g}_2,
\end{equation}
where $\mathfrak{g}_1 = \mathrm{span}\{x_1, x_2, x_{3}\}$ and $\mathfrak{g}_2 = \mathrm{span}\{x_{4}, x_{5}, x_{6}\}$.

\begin{figure}[!htp]
    \centering
    \includegraphics[width=1\linewidth]{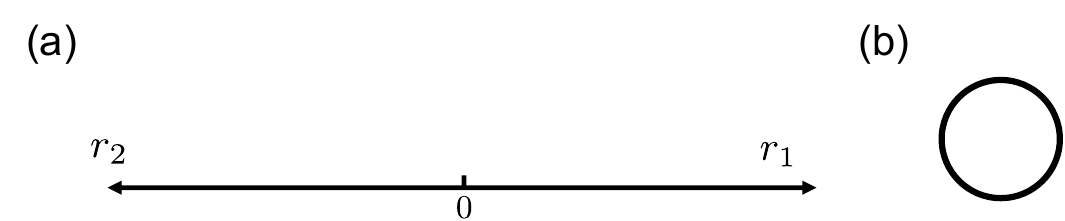}
    \caption{(a) Root system and (b) Dynkin diagram of $\mathfrak{g}_1$.}
    \label{fig:root_system_and_Dynkin_diagram_Hubbard_model}
\end{figure}

For the $3$-dimensional ideal $\mathfrak{g}_1$, the Killing form is non-degenerate, so $\mathfrak{g}_1$ is semisimple. Choosing the Cartan subalgebra $\mathfrak{h} = \mathrm{span}\{x_1\}$ and diagonalizing the adjoint operator $\mathrm{ad}_{x_1}$, we extract two roots, $\{r_1=2x_1, r_2=-2x_1\}$, shown in Fig.~\ref{fig:root_system_and_Dynkin_diagram_Hubbard_model}(a). Taking $\Delta = \{r_1\}$ as the set of simple roots, we arrive at the one-node Dynkin diagram in Fig.~\ref{fig:root_system_and_Dynkin_diagram_Hubbard_model}(b). Therefore,
\begin{equation}
    \mathfrak{g}_1 \cong \mathfrak{su}(2).
\end{equation}

By the same analysis, $\mathfrak{g}_2 \cong \mathfrak{su}(2)$. To make the physical content explicit, we rewrite the generators $x_1,\ldots,x_6$ in the original complex fermion basis:
\begin{gather}
    x_{1}\sim\sum_{\tau=A,B}\begin{pmatrix}c_{\tau,\uparrow}^{\dagger}, & c_{\tau,\downarrow}^{\dagger}\end{pmatrix}\begin{bmatrix}0 & -\i\\
\i & 0
\end{bmatrix}\begin{pmatrix}c_{\tau,\uparrow}\\
c_{\tau,\downarrow}
\end{pmatrix},\\x_{2}\sim\sum_{\tau=A,B}\begin{pmatrix}c_{\tau,\uparrow}^{\dagger}, & c_{\tau,\downarrow}^{\dagger}\end{pmatrix}\begin{bmatrix}1 & 0\\
0 & -1
\end{bmatrix}\begin{pmatrix}c_{\tau,\uparrow}\\
c_{\tau,\downarrow}
\end{pmatrix},\\x_{3}\sim\sum_{\tau=A,B}\begin{pmatrix}c_{\tau,\uparrow}^{\dagger}, & c_{\tau,\downarrow}^{\dagger}\end{pmatrix}\begin{bmatrix}0 & 1\\
1 & 0
\end{bmatrix}\begin{pmatrix}c_{\tau,\uparrow}\\
c_{\tau,\downarrow}
\end{pmatrix},\\x_{4}\sim\sum_{\tau=A,B}\begin{pmatrix}c_{\tau,\uparrow}^{\dagger} & c_{\tau,\downarrow}\end{pmatrix}\begin{bmatrix}1 & 0\\
0 & -1
\end{bmatrix}\begin{pmatrix}c_{\tau,\uparrow}\\
c_{\tau,\downarrow}^{\dagger}
\end{pmatrix},\\x_{5}\sim\sum_{\tau=A,B}\epsilon_{\tau}\begin{pmatrix}c_{\tau,\uparrow}^{\dagger} & c_{\tau,\downarrow}\end{pmatrix}\begin{bmatrix}0 & -\i\\
\i & 0
\end{bmatrix}\begin{pmatrix}c_{\tau,\uparrow}\\
c_{\tau,\downarrow}^{\dagger}
\end{pmatrix},\\x_{6}\sim\sum_{\tau=A,B}\epsilon_{\tau}\begin{pmatrix}c_{\tau,\uparrow}^{\dagger} & c_{\tau,\downarrow}\end{pmatrix}\begin{bmatrix}0 & 1\\
1 & 0
\end{bmatrix}\begin{pmatrix}c_{\tau,\uparrow}\\
c_{\tau,\downarrow}^{\dagger}
\end{pmatrix},
\end{gather}
where $\epsilon_\tau=\pm 1$ for $\tau=A,B$, labeling the two sites on a bond. In this form, $\mathfrak{g}_1$ is the total-spin $\mathrm{SU}(2)$ algebra, while $\mathfrak{g}_2$ is the $\eta$-pairing pseudospin $\mathrm{SU}(2)$ algebra. The full continuous symmetry algebra is therefore $\mathfrak{su}(2)\oplus\mathfrak{su}(2)$, with simply connected group $\mathrm{SU}(2)_S\times \mathrm{SU}(2)_\eta$. To determine the faithful symmetry group, we examine which central elements act trivially on the physical Hilbert space. Each $\mathrm{SU}(2)$ has center $\{I,-I\}$, where $-I$ corresponds to a $2\pi$ rotation; for convenience we choose the $z$ axis. In the spin sector,
\begin{equation}
    -I_s=e^{i2\pi S_z}=e^{i\pi\hat{N}_{\mathrm{tot}}},
\end{equation}
while in the $\eta$-pairing sector,
\begin{equation}
    -I_{\eta}=e^{i2\pi\eta_z}=e^{i\pi(\hat{N}_{\mathrm{tot}}-V)},
\end{equation}
where $\hat{N}_{\mathrm{tot}}$ is the total particle-number operator and $V$ is the total number of lattice sites. For bipartite lattices with an even number of sites, the diagonal element $(-I,-I)$ acts as $e^{i\pi(2\hat{N}_{\mathrm{tot}}-V)}=I$ on the physical Hilbert space. The faithful symmetry group is therefore
\begin{equation}
    \mathrm{SO}(4)\cong \mathrm{SU}(2)_S\times \mathrm{SU}(2)_\eta/\mathbb{Z}_2,
\end{equation}
where $\mathbb{Z}_2=\{(I,I),(-I,-I)\}$. This reproduces the well-known $\mathrm{SO}(4)$ symmetry of the Hubbard model on bipartite lattices~\cite{yang_$eta$_1989,yang_$so_4$_1990}. On a non-bipartite lattice, by contrast, geometrical frustration reduces the $\eta$-pairing $\mathrm{SU}(2)$ symmetry to the $\mathrm{U}(1)$ phase symmetry generated by $x_4$, namely total charge conservation~\cite{moudgalya_symmetries_2023}.

\subsubsection{Bilayer Spin-1/2 Model}

We next consider an AA-stacked bilayer of spin-$1/2$ fermions on the honeycomb lattice. The Hamiltonian is
\begin{gather} \label{eq:bilayer-Hamiltonian}
    H=H_{0}+H_{I},\\H_{0}=-t\sum_{\left\langle i,j\right\rangle ,\lambda,\sigma}c_{i,\lambda,\sigma}^{\dagger}c_{j,\lambda,\sigma}+\mathrm{h.c.,}\\H_{I}=J\sum_{i}\left[\vec{S}_{i,1}\cdot\vec{S}_{i,2}+\frac{1}{4}\left(\rho_{i,1}\rho_{i,2}-1\right)\right],
\end{gather}
where $\lambda\in\{1,2\}$ denotes the layer index, $\vec{S}_{i,\lambda}$ represents the spin operator on layer $\lambda$ at site $i$, and $\rho_{i,\lambda} = n_{i,\lambda} - 1$ measures the local density deviation from half-filling. The corresponding local Hamiltonian $H_b$, including both hopping and interaction terms, is
\begin{equation}
    \begin{aligned}
        H_{b}=&-t\left(c_{i}^{\dagger}c_{j}+c_{j}^{\dagger}c_{i}\right)+\frac{J}{z}\left[\vec{S}_{i,1}\cdot\vec{S}_{i,2}+\frac{1}{4}\left(\rho_{i,1}\rho_{i,2}-1\right)\right.\\&\left.+\vec{S}_{j,1}\cdot\vec{S}_{j,2}+\frac{1}{4}\left(\rho_{j,1}\rho_{j,2}-1\right)\right].
    \end{aligned}
\end{equation}
The local degrees of freedom now consist of the two sites on a bond together with the particle-hole, layer, and spin indices:
\begin{equation} \label{eq:dof}
    \underset{\text{bond}}{\begin{bmatrix}A\\
B
\end{bmatrix}}\otimes\underset{\text{particle-hole}}{\begin{bmatrix}c^{\dagger}\\
c
\end{bmatrix}}\otimes\underset{\text{layer}}{\begin{bmatrix}1\\
2
\end{bmatrix}}\otimes\underset{\text{spin}}{\begin{bmatrix}\uparrow\\
\downarrow
\end{bmatrix}}.
\end{equation}
The local Majorana space is therefore $2 \Nf=2^4=16$ dimensional.

As before, we use tensor products of Pauli matrices to construct a basis for $\mathfrak{so}(16)$:
$\sigma(i,j,k,l) = \sigma(i) \otimes \sigma(j) \otimes \sigma(k) \otimes \sigma(l)$, where the tensor-product order follows Eq.~\eqref{eq:dof}: (bond) $\otimes$ (Majorana) $\otimes$ (layer) $\otimes$ (spin). Antisymmetry requires an odd number of $\sigma(2)$ factors in each tensor product. The basis of $\mathfrak{so}(16)$ therefore contains $120$ matrices with either one or three indices equal to $2$.

Expressing the local Hamiltonian $H_b$ above in the Majorana basis and solving Eq.~\eqref{eq:commutation-condition}, we obtain an $11$-dimensional Lie algebra $\mathfrak{g}^{\prime}$ with basis
\begin{equation} \label{eq:bilayer-lie-algebra-basis}
    \begin{aligned}
        \begin{cases}
            y_{1}=\sigma\left(3,1,2,0\right)\\
            y_{2}=\sigma\left(3,1,2,1\right)\\
            y_{3}=\sigma\left(3,1,2,3\right)\\
            y_{4}=\sigma\left(0,2,0,0\right)\\
            y_{5}=\sigma\left(0,2,0,1\right)
            \end{cases},&\begin{cases}
            y_{6}=\sigma\left(0,2,0,3\right)\\
            y_{7}=\sigma\left(3,3,2,0\right)\\
            y_{8}=\sigma\left(3,3,2,1\right)\\
            y_{9}=\sigma\left(3,3,2,3\right)\\
            y_{10}=\sigma\left(0,0,0,2\right)
            \end{cases},\\y_{11}=\sigma\left(0,2,3,0\right).&
    \end{aligned}
\end{equation}
These basis matrices are orthonormal under the normalized inner product $\langle y_i,y_j \rangle = \frac{1}{16}\mathrm{Tr}(y_i^{\dagger}y_j)$. Computing the commutators $[y_i, y_j]$ shows that the algebra decomposes into a direct sum of two ideals,
\begin{equation}
    \mathfrak{g}^{\prime} = \mathfrak{g}_1^{\prime} \oplus \mathfrak{g}_2^{\prime},
\end{equation}
where $\mathfrak{g}_1^{\prime} = \mathrm{span}\{y_1, \dots, y_{10}\}$ and $\mathfrak{g}_2^{\prime} = \mathrm{span}\{y_{11}\}$. Evidently, the latter generates a $\mathrm{U}(1)$ continuous symmetry.

\begin{figure}[!htp]
    \centering
    \includegraphics[width=1\linewidth]{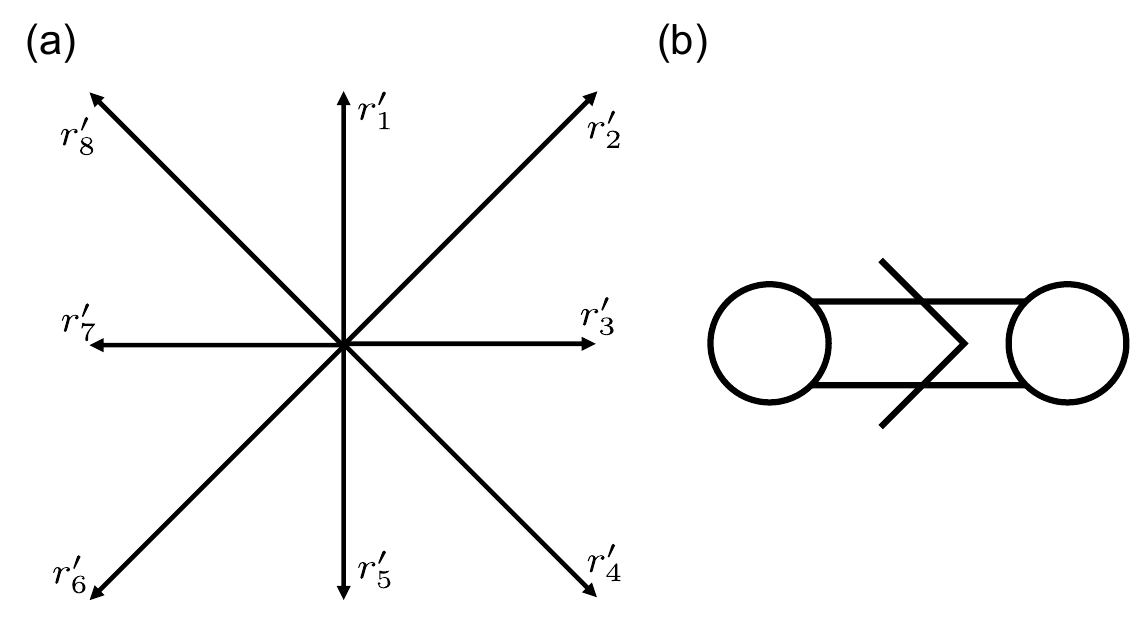}
    \caption{(a) Root system and (b) Dynkin diagram of $\mathfrak{g}_1^{\prime}$.}
    \label{fig:root_system_and_Dynkin_diagram_bilayer_model}
\end{figure}

For the $10$-dimensional ideal $\mathfrak{g}_1^{\prime}$, the Killing form is non-degenerate, so $\mathfrak{g}_1^{\prime}$ is semisimple. We choose a Cartan subalgebra $\mathfrak{h}^{\prime} = \mathrm{span}\{y_1, y_2\}$ and simultaneously diagonalize the adjoint operators $\mathrm{ad}_{y_1}$ and $\mathrm{ad}_{y_2}$. The resulting root system contains eight roots $\{r_i^{\prime}\}$ and is shown in Fig.~\ref{fig:root_system_and_Dynkin_diagram_bilayer_model}(a). A convenient choice of simple roots is $\Delta^{\prime} = \{r_1^{\prime}, r_4^{\prime}\}$:
\begin{equation}
    r_1^{\prime} = 2y_2, \quad r_4^{\prime} = 2y_1 - 2y_2.
\end{equation}
The corresponding Dynkin diagram has two nodes connected by a double edge, with an arrow pointing from the longer root ($r_4^\prime$) to the shorter one ($r_1^\prime$); see Fig.~\ref{fig:root_system_and_Dynkin_diagram_bilayer_model}(b). We therefore identify
\begin{equation}
    \mathfrak{g}_1^{\prime} \cong \mathfrak{so}(5).
\end{equation}

On the honeycomb lattice, every nearest-neighbor bond connects two different sublattices, so no geometrical frustration arises when extending the local symmetry to the full lattice. The symmetry algebra of the total Hamiltonian is therefore $\mathfrak{so}(5)\oplus\mathfrak{u}(1)$. By the same reasoning as above, the faithful symmetry group acting on the Fock space is
\begin{equation}
    G=\bigl[\mathrm{Spin}(5)\times \mathrm{U}(1)\bigr]/\mathbb{Z}_2,
\end{equation}
where $\mathbb{Z}_2=\{(I,I),(-I,-I)\}$. The nontrivial central element of $\mathrm{Spin}(5)$ is $-I \sim e^{i\pi y_6/8}=e^{i\pi S_z^{\mathrm{tot}}}=e^{i\pi\hat N}$, which corresponds to a $2\pi$ rotation. The nontrivial order-two element of $\mathrm{U}(1)$ is $-I \sim e^{i\pi y_{11}/8}=e^{i\pi(\hat N_1-\hat N_2)}=e^{i\pi\hat N}$. Hence the diagonal $\mathbb{Z}_2$ acts trivially on the Fock space, giving the quotient above.

\section{Identification of Candidate Order Parameters}
\label{sec:order_parameter_search}

In Landau theory, a phase transition is characterized by an order parameter transforming in a non-trivial irrep of the symmetry group $G$. If the order parameter transformed in a reducible representation, the free energy would generally split into independent sectors. Except at finely tuned multicritical points, this would correspond to separate phase transitions rather than a single one. The search for candidate order parameters is therefore equivalent to identifying the non-trivial irreps of $G$ within the space of physical operators.

\begin{figure*}[htbp]
    \centering
    \begin{tikzpicture}[
  every node/.style={font=\footnotesize},
  startstop/.style={rectangle, rounded corners, minimum width=3cm, text width=8cm, text centered, inner sep=1.2ex, draw, thick},
  process/.style={rectangle, minimum width=3cm, text width=8cm, text centered, inner sep=1.2ex, draw},
  process_small/.style={rectangle, minimum width=2.5cm, text width=3.2cm, text centered, inner sep=1ex, draw},
  decision/.style={diamond, aspect=4.0, text width=3.8cm, text centered, inner sep=0.2ex, draw},
  arrow/.style={-{LaTeX}, thick},
  every edge quotes/.style={font=\footnotesize, auto}
]

  \node[startstop] (start) at (0, 0) {Determine Majorana degrees of freedom ($2N$) and fundamental representation $\phi$};
  \node[process] (tensor) at (0, -1.3) {Construct induced exterior-power representation $\phi^{(m)}$ for $m$-fermion order parameters};
  \node[process] (intertwiner) at (0, -2.5) {Solve for the complete intertwiner space $C_{\phi^{(m)}}(\mathfrak{g})$};

  \node[decision] (is_irrep) at (0, -4.0) {Is $\dim C_{\phi^{(m)}} = 1$?};

  \node[process_small] (complete) at (0, -5.8) {Current representation decomposition complete};

  \node[decision] (all_irrep) at (0, -7.6) {Are all sub-representations irreducible?};

  \node[process] (combine) at (0, -9.4) {Combine transformation matrices};
  \node[process] (extract) at (0, -10.4) {Extract irreducible invariant subspaces};
  \node[startstop] (end) at (0, -11.4) {Construct physical candidate order parameters};

  \node[process_small] (operator) at (5.5, -4.0) {Construct generic intertwiner $\rho^*$};
  \node[process_small] (eigenspace) at (5.5, -5.3) {Solve for eigenspaces of $\rho^*$};
  \node[process_small] (decompose) at (5.5, -6.6) {Block-diagonalize representation $\phi^{(m)}$};

  \draw[arrow] (start) -- (tensor);
  \draw[arrow] (tensor) -- (intertwiner);
  \draw[arrow] (intertwiner) -- (is_irrep);

  \draw[arrow] (is_irrep.south) -- node[left] {Yes} (complete.north);

  \draw[arrow] (is_irrep.east) -- node[above] {No} (operator.west);

  \draw[arrow] (operator) -- (eigenspace);
  \draw[arrow] (eigenspace) -- (decompose);
  \draw[arrow] (decompose.west) -- ++(-0.8, 0) |- (complete.east);

  \draw[arrow] (complete) -- (all_irrep);

  \draw[arrow] (all_irrep.south) -- node[left] {Yes} (combine.north);

  \draw[arrow] (all_irrep.west) -- node[above, near start] {No} ++(-2.0,0) |- node[left, text width=2.6cm, align=right, pos=0.25] {\footnotesize Recursively decompose each reducible sub-representation} (intertwiner.west);

  \draw[arrow] (combine) -- (extract);
  \draw[arrow] (extract) -- (end);

\end{tikzpicture}
    \caption{Flowchart illustrating the systematic decomposition of the induced exterior-power representation on the $m$-fermion operator space to identify candidate physical order parameters in interacting systems.}
    \label{fig:order_parameter_flowchart}
\end{figure*}
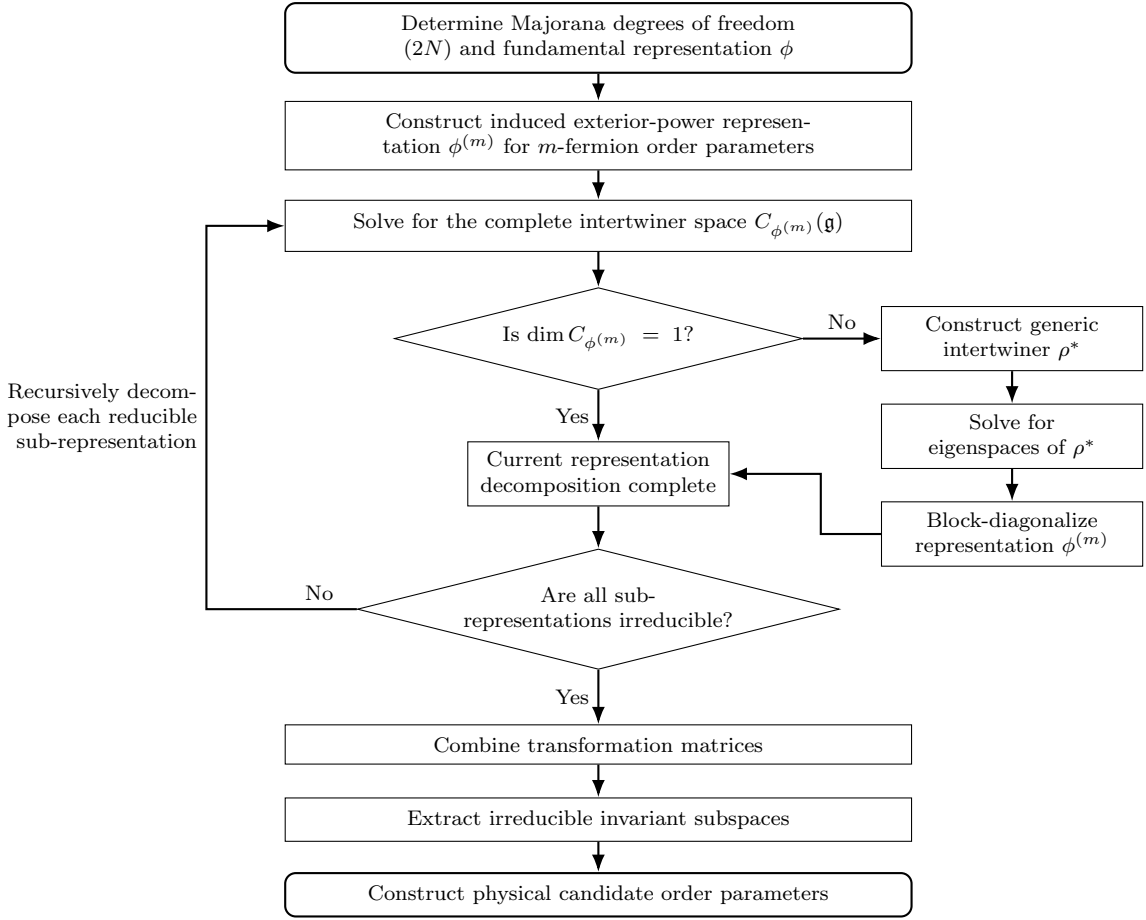

For a connected Lie group, a group representation is irreducible if and only if the corresponding Lie-algebra representation is irreducible. The problem of finding irreps of the continuous symmetry group thus reduces to decomposing the corresponding Lie-algebra representations, and then incorporating the discrete symmetries that connect different connected components of the full symmetry group. Figure~\ref{fig:order_parameter_flowchart} summarizes the overall procedure for extracting candidate order parameters from local degrees of freedom. The subsequent subsections spell out the algorithmic steps in detail.

\subsection{Irreducible Decomposition Algorithm}

Let $V$ be the $2\Nf$-dimensional Majorana vector space. The space of $m$-fermion operators is naturally identified with $\bigwedge^m V$ and transforms under the induced $m$th exterior-power representation $\phi^{(m)} = \bigwedge^m \phi$, where $\phi$ denotes the fundamental representation of the Lie algebra $\mathfrak{g}$ on $V$. Explicitly,
\begin{equation}
    \phi^{(m)}(X)(v_1 \wedge \cdots \wedge v_m)
    =
    \sum_{k=1}^m
    v_1 \wedge \cdots \wedge \phi(X)v_k \wedge \cdots \wedge v_m,
\end{equation}
for $X \in \mathfrak{g}$ and $v_1,\dots,v_m \in V$.

To decompose this generally reducible representation into irreducible components, we calculate its intertwiner space, namely the space of linear endomorphisms that commute with the Lie algebra action:
\begin{equation}
    C_{\phi^{(m)}}(\mathfrak{g}) = \left\{ \rho \in \mathrm{End}(\bigwedge^m V) \mid [\rho, \phi^{(m)}(X)] = 0, \forall X \in \mathfrak{g} \right\}.
\end{equation}
By Schur's lemma, a representation is irreducible if and only if its intertwiner space is trivial, meaning that it consists only of scalar multiples of the identity. Conversely, non-trivial intertwiners provide a systematic way to block-diagonalize and decompose the representation.

Our iterative decomposition algorithm proceeds as follows:
\begin{enumerate}
    \item \textbf{Representation Setup}: Construct a complete basis for the $m$-fermion operator space and define the induced exterior-power representation $\phi^{(m)}$ explicitly.
    \item \textbf{Intertwiner Computation}: Impose the linear constraints $[\rho, \phi^{(m)}(X_a)] = 0$ for all basis generators $X_a \in \mathfrak{g}$. Solving the resulting homogeneous system yields a complete basis for the intertwiner space $C_{\phi^{(m)}}(\mathfrak{g})$.
    \item \textbf{Reducibility Criterion}: If $\dim C_{\phi^{(m)}} = 1$, then the representation is irreducible and the algorithm terminates for this sector.
    \item \textbf{Invariant Subspace Extraction}: If $\dim C_{\phi^{(m)}} > 1$, form a generic intertwiner $\rho^*$ as a random linear combination of the basis elements of $C_{\phi^{(m)}}(\mathfrak{g})$. This probabilistic construction practically guarantees that $\rho^*$ has at least two distinct eigenvalues. Its eigenspaces $V_\lambda$ are then invariant under the action of $\phi^{(m)}$.
    \item \textbf{Recursive Decomposition}: Project $\phi^{(m)}$ onto each eigenspace $V_\lambda$, thereby block-diagonalizing the representation. Apply the same procedure recursively to each block until all subspaces are irreducible.
\end{enumerate}

\subsection{Discrete Symmetries and Isomorphic Representations}

The decomposition algorithm above classifies operators only with respect to the connected component of the continuous symmetry group generated by the Lie algebra $\mathfrak{g}$. To obtain the true irreps of the full physical symmetry group, one must also account for discrete symmetry transformations that bridge disjoint components. In crystalline systems, these include discrete spatial operations such as point-group symmetries, lattice translations, and time-reversal symmetry, all of which constrain the allowed physical order parameters.

When a discrete symmetry operation, such as a sublattice or layer exchange, maps one $\mathfrak{g}$-invariant subspace to another, two cases can arise. If the two subspaces are non-isomorphic as representations of $\mathfrak{g}$, then their direct sum forms an irrep of the full symmetry group. If they are isomorphic, their direct sum is generally reducible once the discrete symmetry is included. Because the two subspaces transform identically under the continuous symmetry, one can further recombine their basis vectors to obtain new invariant subspaces, which then furnish the proper irreps of the full group.

Because these two cases lead to different physical classifications, deciding whether two subspaces are isomorphic is essential. Two representations $\phi_1$ and $\phi_2$ on spaces $V_1$ and $V_2$ are isomorphic if and only if they are related by a bijective intertwiner. We show this algorithmically as follows:
\begin{enumerate}
    \item \textbf{Dimensional Constraint Check}: Compare $\dim V_1$ and $\dim V_2$. If the dimensions differ, the representations cannot be isomorphic.
    \item \textbf{Intertwiner Mapping Evaluation}: If the dimensions agree, compute the space of intertwiners
    $C_{\phi_{1},\phi_{2}}(\mathfrak{g}) = \{ \rho \in \mathrm{Hom}(V_{1},V_{2}) \mid \rho \circ \phi_{1}(X) = \phi_{2}(X) \circ \rho, \forall X \in \mathfrak{g} \}$.
    Solving these linear constraints over a basis of $\mathfrak{g}$ yields a spanning set $\{\rho_i\}$ for $C_{\phi_{1},\phi_{2}}(\mathfrak{g})$.
    \item \textbf{Invertibility Validation}: Form a random linear combination $\rho^{*} = \sum_{i} \alpha_{i} \rho_{i}$ with coefficients $\alpha_i$ sampled uniformly from $(0,1)$, and evaluate $\det(\rho^{*})$.
    \begin{itemize}
        \item If $\det(\rho^{*}) \neq 0$, then $\phi_1$ and $\phi_2$ are isomorphic, and $\rho^*$ provides the explicit isomorphism.
        \item If $\det(\rho^{*}) = 0$, then the representations are non-isomorphic.
    \end{itemize}
    This probabilistic approach is robust because the determinant is a polynomial in the parameters $\alpha_i$. Unless that polynomial vanishes identically, its zero set has Lebesgue measure zero, so a random draw yields $\det(\rho^*)=0$ only with vanishing probability.
\end{enumerate}
\subsection{Application}
\label{sec:application_op}

We now apply this framework to the Hubbard model and the bilayer spin-$1/2$ model with Heisenberg exchange and density-density interlayer couplings on the honeycomb lattice. In both cases, we restrict analysis to the degrees of freedom appearing in the local Hamiltonian. On the honeycomb lattice, the bond degrees of freedom reduce naturally to the two bipartite sublattices, denoted $A$ and $B$.

\subsubsection{Hubbard Model on Honeycomb Lattice}

For the Hubbard model on the honeycomb lattice, the relevant Lie algebra is given in Eq.~\eqref{eq:Hubbard-lie-algebra-basis}. The local degrees of freedom are
\begin{equation}
    \underset{\text{sublattice}}{\begin{bmatrix}A\\ B \end{bmatrix}}
    \otimes
    \underset{\text{particle-hole}}{\begin{bmatrix}c^{\dagger}\\ c \end{bmatrix}}
    \otimes
    \underset{\text{spin}}{\begin{bmatrix}\uparrow\\ \downarrow \end{bmatrix}}.
\end{equation}
The continuous symmetry Lie algebra $\mathfrak{g} \cong \mathfrak{su}(2) \oplus \mathfrak{su}(2)$ acts on this vector space. Restricting the search to the bilinear operator subspace of dimension $\binom{8}{2}=28$, we decompose the representation exactly into irreducible components. Classifying the resulting invariant subspaces by the symmetries they break, namely the continuous symmetry $\mathrm{SO}(4)$ and the sublattice exchange symmetry $Z_2^s$, yields seven distinct candidate order parameters.

Table~\ref{tab:order_parameters_Hubbard_model} summarizes these candidate order parameters, labeled by $\tilde{A}_i$. Their explicit matrix forms are listed in Appendix~\ref{sec:order_parameters_Hubbard_model}.

\begin{table}[htbp]
\caption{Classification of bilinear candidate order parameters for the Hubbard model. The table lists the dimension of the irrep (dim) and the symmetries broken by the order parameter. ``Cont.'' refers to the continuous symmetry with Lie algebra $\mathfrak{su}(2) \oplus \mathfrak{su}(2)$.}
\label{tab:order_parameters_Hubbard_model}
\begin{tabular*}{0.8\linewidth}{@{\extracolsep{\fill}}ccc}
\toprule
Broken Sym. & Label & Dim \\
\midrule
\multirow{2}{*}{Cont.} 
 & $\tilde{A}_{1}, \tilde{A}_{2}, \tilde{A}_{3}, \tilde{A}_{5}$ & 3 \\
 & $\tilde{A}_{7}$ & 9 \\
\midrule
Cont. $+ Z_2^s$ & $\tilde{A}_{4}, \tilde{A}_{6}$ & 3 \\
\bottomrule
\end{tabular*}
\end{table}

In the repulsive case ($U>0$), the system undergoes the well-known quantum phase transition from a gapless Dirac semimetal to an antiferromagnetic insulator once $U$ exceeds a critical value~\cite{wang_resolving_2026}. The corresponding order parameter is $\sum_{u}\tilde{A}_{4}\left(u\right)$, summed over unit cells $u$, where the local vector $\tilde{A}_{4}$ is
\begin{equation}
    \begin{aligned}
           \tilde{A}_{4} &= \begin{pmatrix}
             i\left[\left(c_{A,\uparrow}^{\dag}c_{A,\downarrow}-c_{B,\uparrow}^{\dag}c_{B,\downarrow}\right)-\left(c_{A,\downarrow}^{\dag}c_{A,\uparrow}-c_{B,\downarrow}^{\dag}c_{B,\uparrow}\right)\right]\\
             \left(c_{A,\uparrow}^{\dag}c_{A,\downarrow}-c_{B,\uparrow}^{\dag}c_{B,\downarrow}\right)+\left(c_{A,\downarrow}^{\dag}c_{A,\uparrow}-c_{B,\downarrow}^{\dag}c_{B,\uparrow}\right)\\
             \left(c_{A,\uparrow}^{\dag}c_{A,\uparrow}-c_{A,\downarrow}^{\dag}c_{A,\downarrow}\right)-\left(c_{B,\uparrow}^{\dag}c_{B,\uparrow}-c_{B,\downarrow}^{\dag}c_{B,\downarrow}\right)
           \end{pmatrix}.
    \end{aligned}
\end{equation} 
In the attractive regime ($U<0$), a partial particle-hole transformation maps the Hamiltonian exactly onto a repulsive Hubbard model with the sign of the interaction reversed. As $|U|$ increases, the ground state then evolves from a Dirac semimetal to a superconducting phase. Under this mapping, the antiferromagnetic order parameter $\sum_{u}\tilde{A}_{4}\left(u\right)$ in the repulsive model is mapped to the superconducting and charge-density-wave order parameter $\sum_{u}\tilde{A}_{2}\left(u\right)$ in the attractive model, where $\tilde{A}_{2}$ is

\begin{equation}
    \begin{aligned}
        \tilde{A}_{2} &= \begin{pmatrix}
          i\left[\left(c_{A,\uparrow}^{\dag}c_{A,\downarrow}^{\dag}+c_{B,\uparrow}^{\dag}c_{B,\downarrow}^{\dag}\right)+\left(c_{A,\uparrow}c_{A,\downarrow}+c_{B,\uparrow}c_{B,\downarrow}\right)\right]\\
          \left(c_{A,\uparrow}^{\dag}c_{A,\downarrow}^{\dag}+c_{B,\uparrow}^{\dag}c_{B,\downarrow}^{\dag}\right)-\left(c_{A,\uparrow}c_{A,\downarrow}+c_{B,\uparrow}c_{B,\downarrow}\right)\\
          \left(c_{A,\uparrow}^{\dag}c_{A,\uparrow}+c_{A,\downarrow}^{\dag}c_{A,\downarrow}\right)-\left(c_{B,\uparrow}^{\dag}c_{B,\uparrow}+c_{B,\downarrow}^{\dag}c_{B,\downarrow}\right)
        \end{pmatrix}.
    \end{aligned}
\end{equation}

\subsubsection{Bilayer Spin-1/2 Model on Honeycomb Lattice}

For the AA-stacked bilayer spin-$1/2$ model, the local basis is enlarged by the layer degree of freedom:
\begin{equation}
    \underset{\text{sublattice}}{\begin{bmatrix}A\\ B \end{bmatrix}}
    \otimes
    \underset{\text{particle-hole}}{\begin{bmatrix}c^{\dagger}\\ c \end{bmatrix}}
    \otimes
    \underset{\text{layer}}{\begin{bmatrix}1\\ 2 \end{bmatrix}}
    \otimes
    \underset{\text{spin}}{\begin{bmatrix}\uparrow\\ \downarrow \end{bmatrix}}.
\end{equation}
The symmetry algebra, previously identified in Eq.~\eqref{eq:bilayer-lie-algebra-basis} as $\mathfrak{g}' \cong \mathfrak{so}(5) \oplus \mathfrak{u}(1)$, acts on this $16$-dimensional space. Restricting attention to bilinear fermion operators, a space of dimension $\binom{16}{2}=120$, we use the intertwiner method to decompose the representation into irreducible components. Classifying those components by the symmetries they break, namely the continuous symmetry $\mathrm{Spin}(5) \times \mathrm{U}(1)/\mathbb{Z}_2$, the layer-exchange symmetry $Z_2^l$, and the sublattice-exchange symmetry $Z_2^s$, gives 18 independent local candidate order parameters.

Table~\ref{tab:order_parameters_bilayer_model} summarizes these candidate order parameters, labeled by $\tilde{B}_i$ and grouped by the symmetries they break. Their explicit expressions are given in Appendix~\ref{sec:order_parameters_bilayer_model}.

\begin{table}[htbp]
\caption{Classification of bilinear candidate order parameters for the bilayer model. The table lists the dimension of the irrep (dim) and the symmetries broken by the order parameter. ``Cont.'' refers to the continuous symmetry with Lie algebra $\mathfrak{so}(5)\oplus\mathfrak{u}(1)$.}
\label{tab:order_parameters_bilayer_model}
\begin{tabular*}{0.9\linewidth}{@{\extracolsep{\fill}}ccc}
\toprule
Broken Sym. & Label & Dim \\
\midrule
\multirow{4}{*}{Cont.} 
 & $\tilde{B}_{2}$ & 2 \\
 & $\tilde{B}_{10}, \tilde{B}_{11}, \tilde{B}_{12}, \tilde{B}_{13}$ & 5 \\
 & $\tilde{B}_{7}, \tilde{B}_{8}, \tilde{B}_{9}, \tilde{B}_{15}, \tilde{B}_{16}, \tilde{B}_{17}, \tilde{B}_{18}$ & 10 \\
 & $\tilde{B}_{14}$ & 20 \\
\midrule
$Z_2^l$ & $\tilde{B}_{5}$ & 1 \\
\midrule
$Z_2^l + Z_2^s$ & $\tilde{B}_{4}, \tilde{B}_{6}$ & 1 \\
\midrule
Cont. $+ Z_2^s$ & $\tilde{B}_{1}, \tilde{B}_{3}$ & 2 \\
\bottomrule
\end{tabular*}
\end{table}

\section{Conclusion}
\label{sec:conclusion}

In this work, we developed an algorithmic framework for continuous symmetry analysis and for enumerating candidate order parameters in interacting fermion systems. In the Majorana representation, the symmetry problem is reduced to determining the Lie algebra commuting with the Hamiltonian as a subalgebra of $\mathfrak{so}(2\Nf)$. We then use Cartan subalgebras, root systems, and Dynkin diagrams to identify its semisimple structure and the corresponding locally isomorphic Lie group, and finally determine the faithful symmetry group from the action of the center on the physical Hilbert space.

We also reformulated the search for physical order parameters as the problem of identifying irreducible representations inside the exterior-power representations induced by the symmetry algebra on the Majorana space. Using intertwiners, we gave a practical decomposition scheme for these representation spaces and then incorporated the effects of discrete symmetries to obtain the physically relevant order-parameter sectors.

We illustrated the method with two models on the honeycomb lattice. For the Hubbard model, we recovered the $\mathrm{SO}(4)$ symmetry and classified all $7$ candidate bilinear order parameters. For the bilayer model with Heisenberg exchange and density-density interlayer couplings, we identified the $\mathrm{Spin}(5) \times \mathrm{U}(1)/\mathbb{Z}_2$ symmetry and obtained a complete classification of $18$ candidate bilinear order parameters.

The same framework has been applied to a closely related bilayer honeycomb model with pure Heisenberg interlayer coupling, whose symmetry algebra enlarges to $3\,\mathfrak{su}(2)$ (with faithful symmetry group $\mathrm{SU}(2)^3/\mathbb{Z}_2$, as determined in~\cite{He2026a}). In the companion paper~\cite{He2026a}, the classification of all $19$ symmetry-inequivalent order parameters plays a crucial role in establishing that none of them develops long-range order in the gapped phase, providing numerically exact evidence for symmetric mass generation. For the $\mathrm{Spin}(5) \times \mathrm{U}(1)/\mathbb{Z}_2$ model studied here, quantum Monte Carlo simulations instead reveal an intermediate excitonic phase between the Dirac semimetal and the SMG phase, and the order parameter of this phase can be identified directly from the classification obtained in this work (see Eq.~\eqref{eq:B-1} in the Appendix).

More broadly, the point of this framework is to replace guesswork with a controlled algebraic procedure. That becomes especially valuable in systems with many internal degrees of freedom and multiple competing orders, where intuitive symmetry analysis quickly becomes unreliable. The method is readily extendable to multi-orbital, valley, and moir\'{e} systems, and its algorithmic structure makes full automation natural. It can also be generalized beyond bilinear order parameters to higher-order operators, opening the way to a systematic classification of composite and multipolar orders in strongly correlated systems.

\begin{acknowledgments}
CHH and XYX are supported by the National Natural Science Foundation of China (Grants No. 12447103, No. 12274289), the National Key R\&D Program of China (Grants No. 2022YFA1402702, No. 2021YFA1401400), the Innovation Program for Quantum Science and Technology (under Grant No. 2021ZD0301902), Yangyang Development Fund, and Shanghai Jiao Tong University 2030 Initiative. YZY is supported by the National Science Foundation Grant No. DMR-2238360.
\end{acknowledgments}

\bibliography{main.bib}

\onecolumngrid
\newpage
\appendix
\setcounter{secnumdepth}{2}

\section{Invariant Subspaces of the Lie Algebras}

Below, we enumerate the invariant subspaces of the underlying Lie algebras for both the Hubbard model and the bilayer spin-$1/2$ model on a honeycomb lattice.

\subsection{Hubbard Model on Honeycomb Lattice} \label{sec:invariant_Hubbard_model}

For the Hubbard model on a honeycomb lattice, the continuous symmetry Lie algebra of the system is $\mathfrak{g}\cong\mathfrak{su}(2)\oplus\mathfrak{su}(2)$. By considering the available local degrees of freedom, which are given by the tensor product $\underset{\text{sublattice}}{\begin{bmatrix}A\\ B \end{bmatrix}}\otimes\underset{\text{particle-hole}}{\begin{bmatrix}c^{\dagger}\\ c \end{bmatrix}}\otimes\underset{\text{spin}}{\begin{bmatrix}\uparrow\\ \downarrow \end{bmatrix}}$, we can systematically deduce the invariant subspaces of this Lie algebra. Their corresponding basis matrices are listed as follows:

\begin{equation} \label{eq:A-1}
    A_{1}=c_{A,\uparrow}^{\dag}c_{B,\uparrow}+c_{A,\downarrow}^{\dag}c_{B,\downarrow}+c_{B,\uparrow}^{\dag}c_{A,\uparrow}+c_{B,\downarrow}^{\dag}c_{A,\downarrow},
\end{equation}

\begin{equation}
    A_{2}=\begin{pmatrix}c_{B,\uparrow}^{\dag}c_{B,\downarrow}^{\dag}\\c_{B,\uparrow}c_{B,\downarrow}\\-\left(c_{B,\uparrow}^{\dag}c_{B,\uparrow}+c_{B,\downarrow}^{\dag}c_{B,\downarrow}-1\right)\end{pmatrix},
\end{equation}

\begin{equation}
    A_{3}=\begin{pmatrix}c_{A,\uparrow}^{\dag}c_{A,\downarrow}^{\dag}\\c_{A,\uparrow}c_{A,\downarrow}\\c_{A,\uparrow}^{\dag}c_{A,\uparrow}+c_{A,\downarrow}^{\dag}c_{A,\downarrow}-1\end{pmatrix},
\end{equation}

\begin{equation}
    A_{4}=\begin{pmatrix}c_{A,\uparrow}^{\dag}c_{B,\downarrow}^{\dag}-c_{A,\downarrow}^{\dag}c_{B,\uparrow}^{\dag}\\c_{A,\uparrow}c_{B,\downarrow}-c_{A,\downarrow}c_{B,\uparrow}\\c_{A,\uparrow}^{\dag}c_{B,\uparrow}+c_{A,\downarrow}^{\dag}c_{B,\downarrow}-c_{B,\uparrow}^{\dag}c_{A,\uparrow}-c_{B,\downarrow}^{\dag}c_{A,\downarrow}\end{pmatrix},
\end{equation}

\begin{equation}
    A_{5}=\begin{pmatrix}c_{B,\uparrow}^{\dag}c_{B,\downarrow}\\c_{B,\downarrow}^{\dag}c_{B,\uparrow}\\c_{B,\uparrow}^{\dag}c_{B,\uparrow}-c_{B,\downarrow}^{\dag}c_{B,\downarrow}\end{pmatrix},
\end{equation}

\begin{equation}
    A_{6}=\begin{pmatrix}c_{A,\uparrow}^{\dag}c_{A,\downarrow}\\c_{A,\downarrow}^{\dag}c_{A,\uparrow}\\c_{A,\uparrow}^{\dag}c_{A,\uparrow}-c_{A,\downarrow}^{\dag}c_{A,\downarrow}\end{pmatrix},
\end{equation}

\begin{equation}
    A_{7}=\begin{pmatrix}c_{A,\uparrow}^{\dag}c_{B,\downarrow}-c_{B,\uparrow}^{\dag}c_{A,\downarrow}\\c_{A,\downarrow}^{\dag}c_{B,\uparrow}-c_{B,\downarrow}^{\dag}c_{A,\uparrow}\\c_{A,\uparrow}^{\dag}c_{B,\uparrow}-c_{A,\downarrow}^{\dag}c_{B,\downarrow}-c_{B,\uparrow}^{\dag}c_{A,\uparrow}+c_{B,\downarrow}^{\dag}c_{A,\downarrow}\end{pmatrix},
\end{equation}

\begin{equation}
    A_{8}=\begin{pmatrix}c_{A,\uparrow}^{\dag}c_{B,\downarrow}+c_{B,\uparrow}^{\dag}c_{A,\downarrow}\\c_{A,\downarrow}^{\dag}c_{B,\uparrow}+c_{B,\downarrow}^{\dag}c_{A,\uparrow}\\c_{A,\uparrow}^{\dag}c_{B,\uparrow}-c_{A,\downarrow}^{\dag}c_{B,\downarrow}+c_{B,\uparrow}^{\dag}c_{A,\uparrow}-c_{B,\downarrow}^{\dag}c_{A,\downarrow}\\c_{A,\uparrow}^{\dag}c_{B,\uparrow}^{\dagger}\\c_{A,\downarrow}^{\dagger}c_{B,\downarrow}^{\dagger}\\c_{A,\uparrow}c_{B,\uparrow}\\c_{A,\downarrow}c_{B,\downarrow}\\c_{A,\uparrow}^{\dag}c_{B,\downarrow}^{\dag}+c_{A,\downarrow}^{\dag}c_{B,\uparrow}^{\dag}\\c_{A,\uparrow}c_{B,\downarrow}+c_{A,\downarrow}c_{B,\uparrow}\end{pmatrix}.
\end{equation}

\subsection{Bilayer Spin-1/2 Model on Honeycomb Lattice}\label{sec:invariant_bilayer_model}

As discussed in the main text, the continuous symmetry Lie algebra of this system is $\mathfrak{g}\cong\mathfrak{so}(5)\oplus\mathfrak{u}(1)$. Given the local degrees of freedom $\underset{\text{sublattice}}{\begin{bmatrix}A\\ B \end{bmatrix}}\otimes\underset{\text{particle-hole}}{\begin{bmatrix}c^{\dagger}\\ c \end{bmatrix}}\otimes\underset{\text{layer}}{\begin{bmatrix}1\\ 2 \end{bmatrix}}\otimes\underset{\text{spin}}{\begin{bmatrix}\uparrow\\ \downarrow \end{bmatrix}}$, we systematically identify the invariant subspaces of the Lie algebra. The results are strictly enumerated as follows:

\begin{equation}
    B_{1}=\left(c_{B,2,\uparrow}^{\dag}c_{B,1,\uparrow}+c_{B,2,\downarrow}^{\dag}c_{B,1,\downarrow}\right),
\end{equation}

\begin{equation}
    B_{2}=\left(c_{A,2,\uparrow}^{\dag}c_{A,1,\uparrow}+c_{A,2,\downarrow}^{\dag}c_{A,1,\downarrow}\right),
\end{equation}

\begin{equation}
    B_{3}=\left(c_{A,2,\uparrow}^{\dag}c_{B,1,\uparrow}+c_{A,2,\downarrow}^{\dag}c_{B,1,\downarrow}-c_{B,2,\uparrow}^{\dag}c_{A,1,\uparrow}-c_{B,2,\downarrow}^{\dag}c_{A,1,\downarrow}\right),
\end{equation}

\begin{equation}
    B_{4}=\left(c_{B,1,\uparrow}^{\dag}c_{B,2,\uparrow}+c_{B,1,\downarrow}^{\dag}c_{B,2,\downarrow}\right),
\end{equation}

\begin{equation}
    B_{5}=\left(c_{A,1,\uparrow}^{\dag}c_{A,2,\uparrow}+c_{A,1,\downarrow}^{\dag}c_{A,2,\downarrow}\right),
\end{equation}

\begin{equation}
    B_{6}=\left(c_{A,1,\uparrow}^{\dag}c_{B,2,\uparrow}+c_{A,1,\downarrow}^{\dag}c_{B,2,\downarrow}-c_{B,1,\uparrow}^{\dag}c_{A,2,\uparrow}-c_{B,1,\downarrow}^{\dag}c_{A,2,\downarrow}\right),
\end{equation}

\begin{equation}
    B_{7}=\left(c_{B,1,\uparrow}^{\dag}c_{B,1,\uparrow}+c_{B,1,\downarrow}^{\dag}c_{B,1,\downarrow}-c_{B,2,\uparrow}^{\dag}c_{B,2,\uparrow}-c_{B,2,\downarrow}^{\dag}c_{B,2,\downarrow}\right),
\end{equation}

\begin{equation}
    B_{8}=\left(c_{A,1,\uparrow}^{\dag}c_{A,1,\uparrow}+c_{A,1,\downarrow}^{\dag}c_{A,1,\downarrow}-c_{A,2,\uparrow}^{\dag}c_{A,2,\uparrow}-c_{A,2,\downarrow}^{\dag}c_{A,2,\downarrow}\right),
\end{equation}

\begin{equation}
    B_{9}=\left(c_{A,1,\uparrow}^{\dag}c_{B,1,\uparrow}+c_{A,1,\downarrow}^{\dag}c_{B,1,\downarrow}+c_{B,2,\uparrow}^{\dag}c_{A,2,\uparrow}+c_{B,2,\downarrow}^{\dag}c_{A,2,\downarrow}\right),
\end{equation}

\begin{equation}
    B_{10}=\left(c_{A,2,\uparrow}^{\dag}c_{B,2,\uparrow}+c_{A,2,\downarrow}^{\dag}c_{B,2,\downarrow}+c_{B,1,\uparrow}^{\dag}c_{A,1,\uparrow}+c_{B,1,\downarrow}^{\dag}c_{A,1,\downarrow}\right),
\end{equation}

\begin{equation}
    B_{11}=\begin{pmatrix}c_{B,2,\uparrow}^{\dag}c_{B,1,\downarrow}\\ c_{B,2,\downarrow}^{\dag}c_{B,1,\uparrow}\\ c_{B,2,\uparrow}^{\dag}c_{B,1,\uparrow}-c_{B,2,\downarrow}^{\dag}c_{B,1,\downarrow}\\ -c_{B,2,\uparrow}^{\dag}c_{B,2,\downarrow}^{\dag}\\ -c_{B,1,\uparrow}c_{B,1,\downarrow} \end{pmatrix},
\end{equation}

\begin{equation}
    B_{12}=\begin{pmatrix}c_{A,2,\uparrow}^{\dag}c_{A,1,\downarrow}\\
c_{A,2,\downarrow}^{\dag}c_{A,1,\uparrow}\\
c_{A,2,\uparrow}^{\dag}c_{A,1,\uparrow}-c_{A,2,\downarrow}^{\dag}c_{A,1,\downarrow}\\
c_{A,2,\uparrow}^{\dag}c_{A,2,\downarrow}^{\dag}\\
c_{A,1,\uparrow}c_{A,1,\downarrow}
\end{pmatrix},
\end{equation}

\begin{equation}
    B_{13}=\begin{pmatrix}c_{A,2,\uparrow}^{\dag}c_{B,1,\downarrow}-c_{B,2,\uparrow}^{\dag}c_{A,1,\downarrow}\\
c_{A,2,\downarrow}^{\dag}c_{B,1,\uparrow}-c_{B,2,\downarrow}^{\dag}c_{A,1,\uparrow}\\
c_{A,2,\uparrow}^{\dagger}c_{B,1,\uparrow}-c_{A,2,\downarrow}^{\dag}c_{B,1,\downarrow}-c_{B,2,\uparrow}^{\dag}c_{A,1,\uparrow}+c_{B,2,\downarrow}^{\dag}c_{A,1,\downarrow}\\
c_{A,2,\uparrow}^{\dag}c_{B,2,\downarrow}^{\dag}-c_{A,2,\downarrow}^{\dag}c_{B,2,\uparrow}^{\dag}\\
c_{A,1,\uparrow}c_{B,1,\downarrow}-c_{A,1,\downarrow}c_{B,1,\uparrow}
\end{pmatrix},
\end{equation}

\begin{equation}
    B_{14}=\begin{pmatrix}c_{B,1,\downarrow}^{\dag}c_{B,2,\uparrow}\\ c_{B,1,\uparrow}^{\dag}c_{B,2,\downarrow}\\ c_{B,1,\uparrow}^{\dag}c_{B,2,\uparrow}-c_{B,1,\downarrow}^{\dag}c_{B,2,\downarrow}\\ -c_{B,2,\uparrow}c_{B,2,\downarrow}\\ -c_{B,1,\uparrow}^{\dag}c_{B,1,\downarrow}^{\dag} \end{pmatrix},
\end{equation}

\begin{equation}
    B_{15}=\begin{pmatrix}c_{A,1,\downarrow}^{\dag}c_{A,2,\uparrow}\\
c_{A,1,\uparrow}^{\dag}c_{A,2,\downarrow}\\
c_{A,1,\uparrow}^{\dag}c_{A,2,\uparrow}-c_{A,1,\downarrow}^{\dag}c_{A,2,\downarrow}\\
c_{A,2,\uparrow}c_{A,2,\downarrow}\\
c_{A,1,\uparrow}^{\dag}c_{A,1,\downarrow}^{\dag}
\end{pmatrix},
\end{equation}

\begin{equation}
    B_{16}=\begin{pmatrix}c_{A,1,\downarrow}^{\dag}c_{B,2,\uparrow}-c_{B,1,\downarrow}^{\dag}c_{A,2,\uparrow}\\
c_{A,1,\uparrow}^{\dag}c_{B,2,\downarrow}-c_{B,1,\uparrow}^{\dag}c_{A,2,\downarrow}\\
c_{A,1,\uparrow}^{\dag}c_{B,2,\uparrow}-c_{A,1,\downarrow}^{\dag}c_{B,2,\downarrow}-c_{B,1,\uparrow}^{\dag}c_{A,2,\uparrow}+c_{B,1,\downarrow}^{\dag}c_{A,2,\downarrow}\\
c_{A,2,\uparrow}c_{B,2,\downarrow}-c_{A,2,\downarrow}c_{B,2,\uparrow}\\
c_{A,1,\uparrow}^{\dag}c_{B,1,\downarrow}^{\dag}-c_{A,1,\downarrow}^{\dag}c_{B,1,\uparrow}^{\dag}
\end{pmatrix},
\end{equation}

\begin{equation}
    B_{17}=\begin{pmatrix}c_{B,1,\uparrow}^{\dag}c_{B,1,\downarrow}-c_{B,2,\uparrow}^{\dag}c_{B,2,\downarrow}\\ c_{B,1,\downarrow}^{\dag}c_{B,1,\uparrow}-c_{B,2,\downarrow}^{\dag}c_{B,2,\uparrow}\\ -\left(c_{B,1,\uparrow}^{\dag}c_{B,2,\downarrow}^{\dag}-c_{B,1,\downarrow}^{\dag}c_{B,2,\uparrow}^{\dag}\right)\\ -\left(c_{B,1,\uparrow}c_{B,2,\downarrow}-c_{B,1,\downarrow}c_{B,2,\uparrow}\right)\\ c_{B,1,\uparrow}^{\dag}c_{B,1,\uparrow}-c_{B,1,\downarrow}^{\dag}c_{B,1,\downarrow}-c_{B,2,\uparrow}^{\dag}c_{B,2,\uparrow}+c_{B,2,\downarrow}^{\dag}c_{B,2,\downarrow} \end{pmatrix},
\end{equation}

\begin{equation}
    B_{18}=\begin{pmatrix}c_{A,1,\uparrow}^{\dag}c_{A,1,\downarrow}-c_{A,2,\uparrow}^{\dag}c_{A,2,\downarrow}\\
c_{A,1,\downarrow}^{\dag}c_{A,1,\uparrow}-c_{A,2,\downarrow}^{\dag}c_{A,2,\uparrow}\\
c_{A,1,\uparrow}^{\dag}c_{A,2,\downarrow}^{\dag}-c_{A,1,\downarrow}^{\dag}c_{A,2,\uparrow}^{\dag}\\
c_{A,1,\uparrow}c_{A,2,\downarrow}-c_{A,1,\downarrow}c_{A,2,\uparrow}\\
c_{A,1,\uparrow}^{\dag}c_{A,1,\uparrow}-c_{A,1,\downarrow}^{\dag}c_{A,1,\downarrow}-c_{A,2,\uparrow}^{\dag}c_{A,2,\uparrow}+c_{A,2,\downarrow}^{\dag}c_{A,2,\downarrow}
\end{pmatrix},
\end{equation}

\begin{equation}
    B_{19}=\begin{pmatrix}c_{A,1,\uparrow}^{\dag}c_{B,1,\downarrow}+c_{B,2,\uparrow}^{\dag}c_{A,2,\downarrow}\\c_{A,1,\downarrow}^{\dag}c_{B,1,\uparrow}+c_{B,2,\downarrow}^{\dag}c_{A,2,\uparrow}\\c_{A,1,\uparrow}^{\dag}c_{B,2,\downarrow}^{\dag}-c_{A,1,\downarrow}^{\dag}c_{B,2,\uparrow}^{\dag}\\-\left(c_{A,2,\uparrow}c_{B,1,\downarrow}-c_{A,2,\downarrow}c_{B,1,\uparrow}\right)\\c_{A,1,\uparrow}^{\dag}c_{B,1,\uparrow}-c_{A,1,\downarrow}^{\dag}c_{B,1,\downarrow}+c_{B,2,\uparrow}^{\dag}c_{A,2,\uparrow}-c_{B,2,\downarrow}^{\dag}c_{A,2,\downarrow}\end{pmatrix},
\end{equation}

\begin{equation}
    B_{20}=\begin{pmatrix}c_{A,2,\uparrow}^{\dag}c_{B,2,\downarrow}+c_{B,1,\uparrow}^{\dag}c_{A,1,\downarrow}\\c_{A,2,\downarrow}^{\dag}c_{B,2,\uparrow}+c_{B,1,\downarrow}^{\dag}c_{A,1,\uparrow}\\-\left(c_{A,2,\uparrow}^{\dag}c_{B,1,\downarrow}^{\dag}-c_{A,2,\downarrow}^{\dag}c_{B,1,\uparrow}^{\dag}\right)\\c_{A,1,\uparrow}c_{B,2,\downarrow}-c_{A,1,\downarrow}c_{B,2,\uparrow}\\c_{A,2,\uparrow}^{\dag}c_{B,2,\uparrow}-c_{A,2,\downarrow}^{\dag}c_{B,2,\downarrow}+c_{B,1,\uparrow}^{\dag}c_{A,1,\uparrow}-c_{B,1,\downarrow}^{\dag}c_{A,1,\downarrow}\end{pmatrix},
\end{equation}

\begin{equation}
    B_{21}=\begin{pmatrix}c_{A,1,\uparrow}^{\dag}c_{B,2,\uparrow}+c_{B,1,\uparrow}^{\dag}c_{A,2,\uparrow}\\
c_{A,1,\downarrow}^{\dag}c_{B,2,\downarrow}+c_{B,1,\downarrow}^{\dag}c_{A,2,\downarrow}\\
c_{A,1,\uparrow}^{\dag}c_{B,2,\downarrow}+c_{B,1,\uparrow}^{\dag}c_{A,2,\downarrow}\\
c_{A,1,\downarrow}^{\dag}c_{B,2,\uparrow}+c_{B,1,\downarrow}^{\dag}c_{A,2,\uparrow}\\
c_{A,1,\uparrow}^{\dag}c_{B,1,\uparrow}^{\dag}\\
c_{A,1,\downarrow}^{\dag}c_{B,1,\downarrow}^{\dag}\\
c_{A,2,\uparrow}c_{B,2,\uparrow}\\
c_{A,2,\downarrow}c_{B,2,\downarrow}\\
c_{A,1,\uparrow}^{\dag}c_{B,1,\downarrow}^{\dag}+c_{A,1,\downarrow}^{\dag}c_{B,1,\uparrow}^{\dag}\\
c_{A,2,\uparrow}c_{B,2,\downarrow}+c_{A,2,\downarrow}c_{B,2,\uparrow}
\end{pmatrix},
\end{equation}

\begin{equation}
    B_{22}=\begin{pmatrix}c_{A,2,\uparrow}^{\dag}c_{B,1,\uparrow}+c_{B,2,\uparrow}^{\dag}c_{A,1,\uparrow}\\
c_{A,2,\downarrow}^{\dag}c_{B,1,\downarrow}+c_{B,2,\downarrow}^{\dag}c_{A,1,\downarrow}\\
c_{A,2,\uparrow}^{\dag}c_{B,1,\downarrow}+c_{B,2,\uparrow}^{\dag}c_{A,1,\downarrow}\\
c_{A,2,\downarrow}^{\dag}c_{B,1,\uparrow}+c_{B,2,\downarrow}^{\dag}c_{A,1,\uparrow}\\
c_{A,2,\uparrow}^{\dag}c_{B,2,\uparrow}^{\dag}\\
c_{A,2,\downarrow}^{\dag}c_{B,2,\downarrow}^{\dag}\\
c_{A,1,\uparrow}c_{B,1,\uparrow}\\
c_{A,1,\downarrow}c_{B,1,\downarrow}\\
c_{A,2,\uparrow}^{\dag}c_{B,2,\downarrow}^{\dag}+c_{A,2,\downarrow}^{\dag}c_{B,2,\uparrow}^{\dag}\\
c_{A,1,\uparrow}c_{B,1,\downarrow}+c_{A,1,\downarrow}c_{B,1,\uparrow}
\end{pmatrix},
\end{equation}

\begin{equation}
    B_{23}=\begin{pmatrix}c_{B,1,\uparrow}^{\dag}c_{B,1,\downarrow}+c_{B,2,\uparrow}^{\dag}c_{B,2,\downarrow}\\c_{B,1,\downarrow}^{\dag}c_{B,1,\uparrow}+c_{B,2,\downarrow}^{\dag}c_{B,2,\uparrow}\\-c_{B,1,\uparrow}^{\dag}c_{B,2,\uparrow}^{\dag}\\-c_{B,1,\downarrow}^{\dag}c_{B,2,\downarrow}^{\dag}\\-c_{B,1,\uparrow}c_{B,2,\uparrow}\\-c_{B,1,\downarrow}c_{B,2,\downarrow}\\-\left(c_{B,1,\uparrow}^{\dag}c_{B,2,\downarrow}^{\dag}+c_{B,1,\downarrow}^{\dag}c_{B,2,\uparrow}^{\dag}\right)\\-\left(c_{B,1,\uparrow}c_{B,2,\downarrow}+c_{B,1,\downarrow}c_{B,2,\uparrow}\right)\\c_{B,1,\uparrow}^{\dag}c_{B,1,\uparrow}+c_{B,2,\uparrow}^{\dag}c_{B,2,\uparrow}-1\\c_{B,1,\downarrow}^{\dag}c_{B,1,\downarrow}+c_{B,2,\downarrow}^{\dag}c_{B,2,\downarrow}-1\end{pmatrix},
\end{equation}

\begin{equation}
    B_{24}=\begin{pmatrix}c_{A,1,\uparrow}^{\dag}c_{A,1,\downarrow}+c_{A,2,\uparrow}^{\dag}c_{A,2,\downarrow}\\
c_{A,1,\downarrow}^{\dag}c_{A,1,\uparrow}+c_{A,2,\downarrow}^{\dag}c_{A,2,\uparrow}\\
c_{A,1,\uparrow}^{\dag}c_{A,2,\uparrow}^{\dag}\\
c_{A,1,\downarrow}^{\dag}c_{A,2,\downarrow}^{\dag}\\
c_{A,1,\uparrow}c_{A,2,\uparrow}\\
c_{A,1,\downarrow}c_{A,2,\downarrow}\\
c_{A,1,\uparrow}^{\dag}c_{A,2,\downarrow}^{\dag}+c_{A,1,\downarrow}^{\dag}c_{A,2,\uparrow}^{\dag}\\
c_{A,1,\uparrow}c_{A,2,\downarrow}+c_{A,1,\downarrow}c_{A,2,\uparrow}\\
c_{A,1,\uparrow}^{\dag}c_{A,1,\uparrow}+c_{A,2,\uparrow}^{\dag}c_{A,2,\uparrow}-1\\
c_{A,1,\downarrow}^{\dag}c_{A,1,\downarrow}+c_{A,2,\downarrow}^{\dag}c_{A,2,\downarrow}-1
\end{pmatrix},
\end{equation}

\begin{equation}
    B_{25}=\begin{pmatrix}-\left(c_{A,2,\uparrow}^{\dag}c_{B,2,\downarrow}-c_{B,1,\uparrow}^{\dag}c_{A,1,\downarrow}\right)\\ -\left(c_{A,2,\downarrow}^{\dag}c_{B,2,\uparrow}-c_{B,1,\downarrow}^{\dag}c_{A,1,\uparrow}\right)\\ c_{A,2,\uparrow}^{\dag}c_{B,1,\uparrow}^{\dag}\\ c_{A,2,\downarrow}^{\dag}c_{B,1,\downarrow}^{\dag}\\ c_{A,1,\uparrow}c_{B,2,\uparrow}\\ c_{A,1,\downarrow}c_{B,2,\downarrow}\\ c_{A,2,\uparrow}^{\dag}c_{B,1,\downarrow}^{\dag}+c_{A,2,\downarrow}^{\dag}c_{B,1,\uparrow}^{\dag}\\ c_{A,1,\uparrow}c_{B,2,\downarrow}+c_{A,1,\downarrow}c_{B,2,\uparrow}\\ -\left(c_{A,2,\uparrow}^{\dag}c_{B,2,\uparrow}-c_{B,1,\uparrow}^{\dag}c_{A,1,\uparrow}\right)\\ -\left(c_{A,2,\downarrow}^{\dag}c_{B,2,\downarrow}-c_{B,1,\downarrow}^{\dag}c_{A,1,\downarrow}\right) \end{pmatrix},
\end{equation}

\begin{equation}
    B_{26}=\begin{pmatrix}c_{A,1,\uparrow}^{\dag}c_{B,1,\downarrow}-c_{B,2,\uparrow}^{\dag}c_{A,2,\downarrow}\\
c_{A,1,\downarrow}^{\dag}c_{B,1,\uparrow}-c_{B,2,\downarrow}^{\dag}c_{A,2,\uparrow}\\
c_{A,1,\uparrow}^{\dag}c_{B,2,\uparrow}^{\dag}\\
c_{A,1,\downarrow}^{\dag}c_{B,2,\downarrow}^{\dag}\\
c_{A,2,\uparrow}c_{B,1,\uparrow}\\
c_{A,2,\downarrow}c_{B,1,\downarrow}\\
c_{A,1,\uparrow}^{\dag}c_{B,2,\downarrow}^{\dag}+c_{A,1,\downarrow}^{\dag}c_{B,2,\uparrow}^{\dag}\\
c_{A,2,\uparrow}c_{B,1,\downarrow}+c_{A,2,\downarrow}c_{B,1,\uparrow}\\
c_{A,1,\uparrow}^{\dag}c_{B,1,\uparrow}-c_{B,2,\uparrow}^{\dag}c_{A,2,\uparrow}\\
c_{A,1,\downarrow}^{\dag}c_{B,1,\downarrow}-c_{B,2,\downarrow}^{\dag}c_{A,2,\downarrow}
\end{pmatrix}.
\end{equation}

\section{Candidate Order Parameters}

In this section, we enumerate all candidate order parameters for the Hubbard model and the bilayer spin-$1/2$ model on a honeycomb lattice, expressed within their respective local bases. All basis elements are constructed to be Hermitian operators.

\subsection{Hubbard Model on Honeycomb Lattice} \label{sec:order_parameters_Hubbard_model}
By combining the invariant subspaces derived in Sec.~\ref{sec:invariant_Hubbard_model} and incorporating the additional sublattice exchange $Z_2^s$ symmetry, we identify the full set of irreducible candidate order parameters, which are presented as follows:

\begin{equation}
    \begin{aligned}
        \tilde{A}_{1}&=A_{3}-A_{2}\\&\sim\begin{pmatrix}i\left[\left(c_{A,\uparrow}^{\dag}c_{A,\downarrow}^{\dag}-c_{B,\uparrow}^{\dag}c_{B,\downarrow}^{\dag}\right)+\left(c_{A,\uparrow}c_{A,\downarrow}-c_{B,\uparrow}c_{B,\downarrow}\right)\right]\\
\left(c_{A,\uparrow}^{\dag}c_{A,\downarrow}^{\dag}-c_{B,\uparrow}^{\dag}c_{B,\downarrow}^{\dag}\right)-\left(c_{A,\uparrow}c_{A,\downarrow}-c_{B,\uparrow}c_{B,\downarrow}\right)\\
\left(c_{A,\uparrow}^{\dag}c_{A,\uparrow}+c_{A,\downarrow}^{\dag}c_{A,\downarrow}-1\right)+\left(c_{B,\uparrow}^{\dag}c_{B,\uparrow}+c_{B,\downarrow}^{\dag}c_{B,\downarrow}-1\right)
\end{pmatrix},
    \end{aligned}
\end{equation}

\begin{equation}
    \begin{aligned}
        \tilde{A}_{2}&=A_{3}+A_{2}\\&\sim\begin{pmatrix}i\left[\left(c_{A,\uparrow}^{\dag}c_{A,\downarrow}^{\dag}+c_{B,\uparrow}^{\dag}c_{B,\downarrow}^{\dag}\right)+\left(c_{A,\uparrow}c_{A,\downarrow}+c_{B,\uparrow}c_{B,\downarrow}\right)\right]\\
\left(c_{A,\uparrow}^{\dag}c_{A,\downarrow}^{\dag}+c_{B,\uparrow}^{\dag}c_{B,\downarrow}^{\dag}\right)-\left(c_{A,\uparrow}c_{A,\downarrow}+c_{B,\uparrow}c_{B,\downarrow}\right)\\
\left(c_{A,\uparrow}^{\dag}c_{A,\uparrow}+c_{A,\downarrow}^{\dag}c_{A,\downarrow}\right)-\left(c_{B,\uparrow}^{\dag}c_{B,\uparrow}+c_{B,\downarrow}^{\dag}c_{B,\downarrow}\right)
\end{pmatrix},
    \end{aligned}
\end{equation}

\begin{equation}
    \begin{aligned}
        \tilde{A}_{3}&=A_{4}\\&\sim\begin{pmatrix}i\left[\left(c_{A,\uparrow}^{\dag}c_{B,\downarrow}^{\dag}-c_{A,\downarrow}^{\dag}c_{B,\uparrow}^{\dag}\right)+\left(c_{A,\uparrow}c_{B,\downarrow}-c_{A,\downarrow}c_{B,\uparrow}\right)\right]\\
\left(c_{A,\uparrow}^{\dag}c_{B,\downarrow}^{\dag}-c_{A,\downarrow}^{\dag}c_{B,\uparrow}^{\dag}\right)-\left(c_{A,\uparrow}c_{B,\downarrow}-c_{A,\downarrow}c_{B,\uparrow}\right)\\
i\left(c_{A,\uparrow}^{\dag}c_{B,\uparrow}+c_{A,\downarrow}^{\dag}c_{B,\downarrow}-c_{B,\uparrow}^{\dag}c_{A,\uparrow}-c_{B,\downarrow}^{\dag}c_{A,\downarrow}\right)
\end{pmatrix},
    \end{aligned}
\end{equation}

\begin{equation}
    \begin{aligned}
            \tilde{A}_{4}&=A_{6}-A_{5}\\&\sim\begin{pmatrix}i\left[\left(c_{A,\uparrow}^{\dag}c_{A,\downarrow}-c_{B,\uparrow}^{\dag}c_{B,\downarrow}\right)-\left(c_{A,\downarrow}^{\dag}c_{A,\uparrow}-c_{B,\downarrow}^{\dag}c_{B,\uparrow}\right)\right]\\
\left(c_{A,\uparrow}^{\dag}c_{A,\downarrow}-c_{B,\uparrow}^{\dag}c_{B,\downarrow}\right)+\left(c_{A,\downarrow}^{\dag}c_{A,\uparrow}-c_{B,\downarrow}^{\dag}c_{B,\uparrow}\right)\\
\left(c_{A,\uparrow}^{\dag}c_{A,\uparrow}-c_{A,\downarrow}^{\dag}c_{A,\downarrow}\right)-\left(c_{B,\uparrow}^{\dag}c_{B,\uparrow}-c_{B,\downarrow}^{\dag}c_{B,\downarrow}\right)
\end{pmatrix},
    \end{aligned}
\end{equation}

\begin{equation}
    \begin{aligned}
            \tilde{A}_{5}&=A_{5}+A_{4}\\&\sim\begin{pmatrix}i\left[\left(c_{A,\uparrow}^{\dag}c_{A,\downarrow}+c_{B,\uparrow}^{\dag}c_{B,\downarrow}\right)-\left(c_{A,\downarrow}^{\dag}c_{A,\uparrow}+c_{B,\downarrow}^{\dag}c_{B,\uparrow}\right)\right]\\
\left(c_{A,\uparrow}^{\dag}c_{A,\downarrow}+c_{B,\uparrow}^{\dag}c_{B,\downarrow}\right)+\left(c_{A,\downarrow}^{\dag}c_{A,\uparrow}+c_{B,\downarrow}^{\dag}c_{B,\uparrow}\right)\\
\left(c_{A,\uparrow}^{\dag}c_{A,\uparrow}-c_{A,\downarrow}^{\dag}c_{A,\downarrow}\right)+\left(c_{B,\uparrow}^{\dag}c_{B,\uparrow}-c_{B,\downarrow}^{\dag}c_{B,\downarrow}\right)
\end{pmatrix},
    \end{aligned}
\end{equation}

\begin{equation}
    \begin{aligned}
        \tilde{A}_{6}&=A_{7}\\&\sim\begin{pmatrix}i\left[\left(c_{A,\uparrow}^{\dag}c_{B,\downarrow}-c_{B,\uparrow}^{\dag}c_{A,\downarrow}\right)+\left(c_{A,\downarrow}^{\dag}c_{B,\uparrow}-c_{B,\downarrow}^{\dag}c_{A,\uparrow}\right)\right]\\
\left(c_{A,\uparrow}^{\dag}c_{B,\downarrow}-c_{B,\uparrow}^{\dag}c_{A,\downarrow}\right)-\left(c_{A,\downarrow}^{\dag}c_{B,\uparrow}-c_{B,\downarrow}^{\dag}c_{A,\uparrow}\right)\\
i\left(c_{A,\uparrow}^{\dag}c_{B,\uparrow}-c_{A,\downarrow}^{\dag}c_{B,\downarrow}-c_{B,\uparrow}^{\dag}c_{A,\uparrow}+c_{B,\downarrow}^{\dag}c_{A,\downarrow}\right)
\end{pmatrix},
    \end{aligned}
\end{equation}

\begin{equation}
    \begin{aligned}
        \tilde{A}_{7}&=A_{8}\\&=\begin{pmatrix}\frac{i}{2}\left[\left(c_{A,\uparrow}^{\dag}c_{B,\downarrow}+c_{B,\uparrow}^{\dag}c_{A,\downarrow}\right)-\left(c_{A,\downarrow}^{\dag}c_{B,\uparrow}+c_{B,\downarrow}^{\dag}c_{A,\uparrow}\right)\right]\\
\frac{1}{2}\left[\left(c_{A,\uparrow}^{\dag}c_{B,\downarrow}+c_{B,\uparrow}^{\dag}c_{A,\downarrow}\right)+\left(c_{A,\downarrow}^{\dag}c_{B,\uparrow}+c_{B,\downarrow}^{\dag}c_{A,\uparrow}\right)\right]\\
\frac{1}{2}\left(c_{A,\uparrow}^{\dag}c_{B,\uparrow}-c_{A,\downarrow}^{\dag}c_{B,\downarrow}+c_{B,\uparrow}^{\dag}c_{A,\uparrow}-c_{B,\downarrow}^{\dag}c_{A,\downarrow}\right)\\
i\left(c_{A,\uparrow}^{\dag}c_{B,\uparrow}^{\dagger}+c_{A,\uparrow}c_{B,\uparrow}\right)\\
i\left(c_{A,\downarrow}^{\dagger}c_{B,\downarrow}^{\dagger}+c_{A,\downarrow}c_{B,\downarrow}\right)\\
c_{A,\uparrow}^{\dag}c_{B,\uparrow}^{\dagger}-c_{A,\uparrow}c_{B,\uparrow}\\
c_{A,\downarrow}^{\dagger}c_{B,\downarrow}^{\dagger}-c_{A,\downarrow}c_{B,\downarrow}\\
\frac{i}{2}\left[\left(c_{A,\uparrow}^{\dag}c_{B,\downarrow}^{\dag}+c_{A,\downarrow}^{\dag}c_{B,\uparrow}^{\dag}\right)+\left(c_{A,\uparrow}c_{B,\downarrow}+c_{A,\downarrow}c_{B,\uparrow}\right)\right]\\
\frac{1}{2}\left[\left(c_{A,\uparrow}^{\dag}c_{B,\downarrow}^{\dag}+c_{A,\downarrow}^{\dag}c_{B,\uparrow}^{\dag}\right)-\left(c_{A,\uparrow}c_{B,\downarrow}+c_{A,\downarrow}c_{B,\uparrow}\right)\right]
\end{pmatrix}.
    \end{aligned}
\end{equation}

Note that the invariant subspace corresponding to Eq.~(\ref{eq:A-1}) constitutes the one-dimensional trivial representation of the full symmetry, and therefore does not represent a symmetry-breaking order parameter.

\subsection{Bilayer Spin-1/2 Model on Honeycomb Lattice} \label{sec:order_parameters_bilayer_model}
By incorporating the invariant subspaces established in Sec.~\ref{sec:invariant_bilayer_model} alongside the discrete layer exchange $Z_2^l$ and sublattice exchange $Z_2^s$ symmetries, we deduce the complete classification of candidate order parameters. They are formulated as follows:

\begin{equation} \label{eq:B-1}
    \begin{aligned}
        \tilde{B}_{1}=&\left(B_{5}-B_{4}\right)\oplus\left(B_{2}-B_{1}\right)\\\sim&\begin{pmatrix}i\begin{bmatrix}\left[\left(c_{A,1,\uparrow}^{\dag}c_{A,2,\uparrow}+c_{A,1,\downarrow}^{\dag}c_{A,2,\downarrow}\right)-\left(c_{B,1,\uparrow}^{\dag}c_{B,2,\uparrow}+c_{B,1,\downarrow}^{\dag}c_{B,2,\downarrow}\right)\right]\\
-\left[\left(c_{A,2,\uparrow}^{\dag}c_{A,1,\uparrow}+c_{A,2,\downarrow}^{\dag}c_{A,1,\downarrow}\right)-\left(c_{B,2,\uparrow}^{\dag}c_{B,1,\uparrow}+c_{B,2,\downarrow}^{\dag}c_{B,1,\downarrow}\right)\right]
\end{bmatrix}\\
\begin{bmatrix}\left(c_{A,1,\uparrow}^{\dag}c_{A,2,\uparrow}+c_{A,1,\downarrow}^{\dag}c_{A,2,\downarrow}\right)-\left(c_{B,1,\uparrow}^{\dag}c_{B,2,\uparrow}+c_{B,1,\downarrow}^{\dag}c_{B,2,\downarrow}\right)\\
+\left(c_{A,2,\uparrow}^{\dag}c_{A,1,\uparrow}+c_{A,2,\downarrow}^{\dag}c_{A,1,\downarrow}\right)-\left(c_{B,2,\uparrow}^{\dag}c_{B,1,\uparrow}+c_{B,2,\downarrow}^{\dag}c_{B,1,\downarrow}\right)
\end{bmatrix}
\end{pmatrix},
    \end{aligned}
\end{equation}

\begin{equation}
    \begin{aligned}
        \tilde{B}_{2}=&\left(B_{5}+B_{4}\right)\oplus\left(B_{2}+B_{1}\right)\\\sim&\begin{pmatrix}i\begin{bmatrix}\left[\left(c_{A,1,\uparrow}^{\dag}c_{A,2,\uparrow}+c_{A,1,\downarrow}^{\dag}c_{A,2,\downarrow}\right)+\left(c_{B,1,\uparrow}^{\dag}c_{B,2,\uparrow}+c_{B,1,\downarrow}^{\dag}c_{B,2,\downarrow}\right)\right]\\
-\left[\left(c_{A,2,\uparrow}^{\dag}c_{A,1,\uparrow}+c_{A,2,\downarrow}^{\dag}c_{A,1,\downarrow}\right)+\left(c_{B,2,\uparrow}^{\dag}c_{B,1,\uparrow}+c_{B,2,\downarrow}^{\dag}c_{B,1,\downarrow}\right)\right]
\end{bmatrix}\\
\begin{bmatrix}\left(c_{A,1,\uparrow}^{\dag}c_{A,2,\uparrow}+c_{A,1,\downarrow}^{\dag}c_{A,2,\downarrow}\right)+\left(c_{B,1,\uparrow}^{\dag}c_{B,2,\uparrow}+c_{B,1,\downarrow}^{\dag}c_{B,2,\downarrow}\right)\\
+\left(c_{A,2,\uparrow}^{\dag}c_{A,1,\uparrow}+c_{A,2,\downarrow}^{\dag}c_{A,1,\downarrow}\right)+\left(c_{B,2,\uparrow}^{\dag}c_{B,1,\uparrow}+c_{B,2,\downarrow}^{\dag}c_{B,1,\downarrow}\right)
\end{bmatrix}
\end{pmatrix},
    \end{aligned}
\end{equation}

\begin{equation}
    \begin{aligned}
        \tilde{B}_{3}=&B_{3}\oplus B_{6}\\\sim&\begin{pmatrix}i\begin{bmatrix}\left(c_{A,2,\uparrow}^{\dag}c_{B,1,\uparrow}+c_{A,2,\downarrow}^{\dag}c_{B,1,\downarrow}-c_{B,2,\uparrow}^{\dag}c_{A,1,\uparrow}-c_{B,2,\downarrow}^{\dag}c_{A,1,\downarrow}\right)\\
+\left(c_{A,1,\uparrow}^{\dag}c_{B,2,\uparrow}+c_{A,1,\downarrow}^{\dag}c_{B,2,\downarrow}-c_{B,1,\uparrow}^{\dag}c_{A,2,\uparrow}-c_{B,1,\downarrow}^{\dag}c_{A,2,\downarrow}\right)
\end{bmatrix}\\
\begin{bmatrix}\left(c_{A,2,\uparrow}^{\dag}c_{B,1,\uparrow}+c_{A,2,\downarrow}^{\dag}c_{B,1,\downarrow}-c_{B,2,\uparrow}^{\dag}c_{A,1,\uparrow}-c_{B,2,\downarrow}^{\dag}c_{A,1,\downarrow}\right)\\
-\left(c_{A,1,\uparrow}^{\dag}c_{B,2,\uparrow}+c_{A,1,\downarrow}^{\dag}c_{B,2,\downarrow}-c_{B,1,\uparrow}^{\dag}c_{A,2,\uparrow}-c_{B,1,\downarrow}^{\dag}c_{A,2,\downarrow}\right)
\end{bmatrix}
\end{pmatrix},
    \end{aligned}
\end{equation}

\begin{equation}
    \begin{aligned}
        \tilde{B}_{4}=&B_{8}-B_{7}\\=&\left(c_{A,1,\uparrow}^{\dag}c_{A,1,\uparrow}+c_{A,1,\downarrow}^{\dag}c_{A,1,\downarrow}-c_{A,2,\uparrow}^{\dag}c_{A,2,\uparrow}-c_{A,2,\downarrow}^{\dag}c_{A,2,\downarrow}\right)\\&-\left(c_{B,1,\uparrow}^{\dag}c_{B,1,\uparrow}+c_{B,1,\downarrow}^{\dag}c_{B,1,\downarrow}-c_{B,2,\uparrow}^{\dag}c_{B,2,\uparrow}-c_{B,2,\downarrow}^{\dag}c_{B,2,\downarrow}\right),
    \end{aligned}
\end{equation}

\begin{equation}
    \begin{aligned}
        \tilde{B}_{5}=&B_{8}+B_{7}\\=&\left(c_{A,1,\uparrow}^{\dag}c_{A,1,\uparrow}+c_{A,1,\downarrow}^{\dag}c_{A,1,\downarrow}-c_{A,2,\uparrow}^{\dag}c_{A,2,\uparrow}-c_{A,2,\downarrow}^{\dag}c_{A,2,\downarrow}\right)\\&+\left(c_{B,1,\uparrow}^{\dag}c_{B,1,\uparrow}+c_{B,1,\downarrow}^{\dag}c_{B,1,\downarrow}-c_{B,2,\uparrow}^{\dag}c_{B,2,\uparrow}-c_{B,2,\downarrow}^{\dag}c_{B,2,\downarrow}\right),
    \end{aligned}
\end{equation}

\begin{equation}
    \begin{aligned}
        \tilde{B}_{6}=&B_{9}-B_{10}\\=&i\left(c_{A,1,\uparrow}^{\dag}c_{B,1,\uparrow}+c_{A,1,\downarrow}^{\dag}c_{B,1,\downarrow}+c_{B,2,\uparrow}^{\dag}c_{A,2,\uparrow}+c_{B,2,\downarrow}^{\dag}c_{A,2,\downarrow}\right)\\&-i\left(c_{A,2,\uparrow}^{\dag}c_{B,2,\uparrow}+c_{A,2,\downarrow}^{\dag}c_{B,2,\downarrow}+c_{B,1,\uparrow}^{\dag}c_{A,1,\uparrow}+c_{B,1,\downarrow}^{\dag}c_{A,1,\downarrow}\right),
    \end{aligned}
\end{equation}

\begin{equation}
    \begin{aligned}
        \tilde{B}_{7}&=\left(B_{15}-B_{14}\right)\oplus\left(B_{12}-B_{11}\right)\\&\sim\begin{pmatrix}i\left[\left(c_{A,2,\uparrow}^{\dag}c_{A,1,\downarrow}-c_{B,2,\uparrow}^{\dag}c_{B,1,\downarrow}\right)-\left(c_{A,1,\downarrow}^{\dag}c_{A,2,\uparrow}-c_{B,1,\downarrow}^{\dag}c_{B,2,\uparrow}\right)\right]\\
i\left[\left(c_{A,2,\downarrow}^{\dag}c_{A,1,\uparrow}-c_{B,2,\downarrow}^{\dag}c_{B,1,\uparrow}\right)-\left(c_{A,1,\uparrow}^{\dag}c_{A,2,\downarrow}-c_{B,1,\uparrow}^{\dag}c_{B,2,\downarrow}\right)\right]\\
\frac{i}{2}\begin{bmatrix}\left[\left(c_{A,2,\uparrow}^{\dag}c_{A,1,\uparrow}-c_{A,2,\downarrow}^{\dag}c_{A,1,\downarrow}\right)-\left(c_{B,2,\uparrow}^{\dag}c_{B,1,\uparrow}-c_{B,2,\downarrow}^{\dag}c_{B,1,\downarrow}\right)\right]\\
-\left[\left(c_{A,1,\uparrow}^{\dag}c_{A,2,\uparrow}-c_{A,1,\downarrow}^{\dag}c_{A,2,\downarrow}\right)-\left(c_{B,1,\uparrow}^{\dag}c_{B,2,\uparrow}-c_{B,1,\downarrow}^{\dag}c_{B,2,\downarrow}\right)\right]
\end{bmatrix}\\
i\left(c_{A,2,\uparrow}^{\dag}c_{A,2,\downarrow}^{\dag}+c_{B,2,\uparrow}^{\dag}c_{B,2,\downarrow}^{\dag}+c_{A,2,\uparrow}c_{A,2,\downarrow}+c_{B,2,\uparrow}c_{B,2,\downarrow}\right)\\
c_{A,1,\uparrow}^{\dag}c_{A,1,\downarrow}^{\dag}+c_{B,1,\uparrow}^{\dag}c_{B,1,\downarrow}^{\dag}-c_{A,1,\uparrow}c_{A,1,\downarrow}-c_{B,1,\uparrow}c_{B,1,\downarrow}\\
\left(c_{A,2,\uparrow}^{\dag}c_{A,1,\downarrow}-c_{B,2,\uparrow}^{\dag}c_{B,1,\downarrow}\right)+\left(c_{A,1,\downarrow}^{\dag}c_{A,2,\uparrow}-c_{B,1,\downarrow}^{\dag}c_{B,2,\uparrow}\right)\\
\left(c_{A,2,\downarrow}^{\dag}c_{A,1,\uparrow}-c_{B,2,\downarrow}^{\dag}c_{B,1,\uparrow}\right)+\left(c_{A,1,\uparrow}^{\dag}c_{A,2,\downarrow}-c_{B,1,\uparrow}^{\dag}c_{B,2,\downarrow}\right)\\
\frac{1}{2}\begin{bmatrix}\left[\left(c_{A,2,\uparrow}^{\dag}c_{A,1,\uparrow}-c_{A,2,\downarrow}^{\dag}c_{A,1,\downarrow}\right)-\left(c_{B,2,\uparrow}^{\dag}c_{B,1,\uparrow}-c_{B,2,\downarrow}^{\dag}c_{B,1,\downarrow}\right)\right]\\
+\left[\left(c_{A,1,\uparrow}^{\dag}c_{A,2,\uparrow}-c_{A,1,\downarrow}^{\dag}c_{A,2,\downarrow}\right)-\left(c_{B,1,\uparrow}^{\dag}c_{B,2,\uparrow}-c_{B,1,\downarrow}^{\dag}c_{B,2,\downarrow}\right)\right]
\end{bmatrix}\\
c_{A,2,\uparrow}^{\dag}c_{A,2,\downarrow}^{\dag}+c_{B,2,\uparrow}^{\dag}c_{B,2,\downarrow}^{\dag}-c_{A,2,\uparrow}c_{A,2,\downarrow}-c_{B,2,\uparrow}c_{B,2,\downarrow}\\
i\left(c_{A,1,\uparrow}^{\dag}c_{A,1,\downarrow}^{\dag}+c_{B,1,\uparrow}^{\dag}c_{B,1,\downarrow}^{\dag}+c_{A,1,\uparrow}c_{A,1,\downarrow}+c_{B,1,\uparrow}c_{B,1,\downarrow}\right)
\end{pmatrix},
    \end{aligned}
\end{equation}

\begin{equation}
    \begin{aligned}
        \tilde{B}_{8}=&\left(B_{15}+B_{14}\right)\oplus\left(B_{12}+B_{11}\right)\\\sim&\begin{pmatrix}i\left[\left(c_{A,2,\uparrow}^{\dag}c_{A,1,\downarrow}+c_{B,2,\uparrow}^{\dag}c_{B,1,\downarrow}\right)-\left(c_{A,1,\downarrow}^{\dag}c_{A,2,\uparrow}+c_{B,1,\downarrow}^{\dag}c_{B,2,\uparrow}\right)\right]\\
i\left[\left(c_{A,2,\downarrow}^{\dag}c_{A,1,\uparrow}+c_{B,2,\downarrow}^{\dag}c_{B,1,\uparrow}\right)-\left(c_{A,1,\uparrow}^{\dag}c_{A,2,\downarrow}+c_{B,1,\uparrow}^{\dag}c_{B,2,\downarrow}\right)\right]\\
\frac{i}{2}\begin{bmatrix}\left[\left(c_{A,2,\uparrow}^{\dag}c_{A,1,\uparrow}-c_{A,2,\downarrow}^{\dag}c_{A,1,\downarrow}\right)+\left(c_{B,2,\uparrow}^{\dag}c_{B,1,\uparrow}-c_{B,2,\downarrow}^{\dag}c_{B,1,\downarrow}\right)\right]\\
-\left[\left(c_{A,1,\uparrow}^{\dag}c_{A,2,\uparrow}-c_{A,1,\downarrow}^{\dag}c_{A,2,\downarrow}\right)+\left(c_{B,1,\uparrow}^{\dag}c_{B,2,\uparrow}-c_{B,1,\downarrow}^{\dag}c_{B,2,\downarrow}\right)\right]
\end{bmatrix}\\
i\left(c_{A,2,\uparrow}^{\dag}c_{A,2,\downarrow}^{\dag}-c_{B,2,\uparrow}^{\dag}c_{B,2,\downarrow}^{\dag}+c_{A,2,\uparrow}c_{A,2,\downarrow}-c_{B,2,\uparrow}c_{B,2,\downarrow}\right)\\
c_{A,1,\uparrow}^{\dag}c_{A,1,\downarrow}^{\dag}-c_{B,1,\uparrow}^{\dag}c_{B,1,\downarrow}^{\dag}-c_{A,1,\uparrow}c_{A,1,\downarrow}+c_{B,1,\uparrow}c_{B,1,\downarrow}\\
\left(c_{A,2,\uparrow}^{\dag}c_{A,1,\downarrow}+c_{B,2,\uparrow}^{\dag}c_{B,1,\downarrow}\right)+\left(c_{A,1,\downarrow}^{\dag}c_{A,2,\uparrow}+c_{B,1,\downarrow}^{\dag}c_{B,2,\uparrow}\right)\\
\left(c_{A,2,\downarrow}^{\dag}c_{A,1,\uparrow}+c_{B,2,\downarrow}^{\dag}c_{B,1,\uparrow}\right)+\left(c_{A,1,\uparrow}^{\dag}c_{A,2,\downarrow}+c_{B,1,\uparrow}^{\dag}c_{B,2,\downarrow}\right)\\
\frac{1}{2}\begin{bmatrix}\left[\left(c_{A,2,\uparrow}^{\dag}c_{A,1,\uparrow}-c_{A,2,\downarrow}^{\dag}c_{A,1,\downarrow}\right)+\left(c_{B,2,\uparrow}^{\dag}c_{B,1,\uparrow}-c_{B,2,\downarrow}^{\dag}c_{B,1,\downarrow}\right)\right]\\
+\left[\left(c_{A,1,\uparrow}^{\dag}c_{A,2,\uparrow}-c_{A,1,\downarrow}^{\dag}c_{A,2,\downarrow}\right)+\left(c_{B,1,\uparrow}^{\dag}c_{B,2,\uparrow}-c_{B,1,\downarrow}^{\dag}c_{B,2,\downarrow}\right)\right]
\end{bmatrix}\\
c_{A,2,\uparrow}^{\dag}c_{A,2,\downarrow}^{\dag}-c_{B,2,\uparrow}^{\dag}c_{B,2,\downarrow}^{\dag}-c_{A,2,\uparrow}c_{A,2,\downarrow}+c_{B,2,\uparrow}c_{B,2,\downarrow}\\
i\left(c_{A,1,\uparrow}^{\dag}c_{A,1,\downarrow}^{\dag}-c_{B,1,\uparrow}^{\dag}c_{B,1,\downarrow}^{\dag}+c_{A,1,\uparrow}c_{A,1,\downarrow}-c_{B,1,\uparrow}c_{B,1,\downarrow}\right)
\end{pmatrix},
    \end{aligned}
\end{equation}

\begin{equation}
    \begin{aligned}
        \tilde{B}_{9}=&B_{16}\oplus B_{13}\\\sim&\begin{pmatrix}i\left[\left(c_{A,1,\downarrow}^{\dag}c_{B,2,\uparrow}-c_{B,1,\downarrow}^{\dag}c_{A,2,\uparrow}\right)+\left(c_{A,2,\uparrow}^{\dag}c_{B,1,\downarrow}-c_{B,2,\uparrow}^{\dag}c_{A,1,\downarrow}\right)\right]\\
i\left[\left(c_{A,1,\uparrow}^{\dag}c_{B,2,\downarrow}-c_{B,1,\uparrow}^{\dag}c_{A,2,\downarrow}\right)+\left(c_{A,2,\downarrow}^{\dag}c_{B,1,\uparrow}-c_{B,2,\downarrow}^{\dag}c_{A,1,\uparrow}\right)\right]\\
\frac{i}{2}\begin{bmatrix}\left(c_{A,1,\uparrow}^{\dag}c_{B,2,\uparrow}-c_{A,1,\downarrow}^{\dag}c_{B,2,\downarrow}-c_{B,1,\uparrow}^{\dag}c_{A,2,\uparrow}+c_{B,1,\downarrow}^{\dag}c_{A,2,\downarrow}\right)\\
+\left(c_{A,2,\uparrow}^{\dagger}c_{B,1,\uparrow}-c_{A,2,\downarrow}^{\dag}c_{B,1,\downarrow}-c_{B,2,\uparrow}^{\dag}c_{A,1,\uparrow}+c_{B,2,\downarrow}^{\dag}c_{A,1,\downarrow}\right)
\end{bmatrix}\\
\left(c_{A,2,\uparrow}^{\dag}c_{B,2,\downarrow}^{\dag}-c_{A,2,\downarrow}^{\dag}c_{B,2,\uparrow}^{\dag}\right)-\left(c_{A,2,\uparrow}c_{B,2,\downarrow}-c_{A,2,\downarrow}c_{B,2,\uparrow}\right)\\
i\left[\left(c_{A,1,\uparrow}^{\dag}c_{B,1,\downarrow}^{\dag}-c_{A,1,\downarrow}^{\dag}c_{B,1,\uparrow}^{\dag}\right)+\left(c_{A,1,\uparrow}c_{B,1,\downarrow}-c_{A,1,\downarrow}c_{B,1,\uparrow}\right)\right]\\
\left(c_{A,1,\downarrow}^{\dag}c_{B,2,\uparrow}-c_{B,1,\downarrow}^{\dag}c_{A,2,\uparrow}\right)-\left(c_{A,2,\uparrow}^{\dag}c_{B,1,\downarrow}-c_{B,2,\uparrow}^{\dag}c_{A,1,\downarrow}\right)\\
\left(c_{A,1,\uparrow}^{\dag}c_{B,2,\downarrow}-c_{B,1,\uparrow}^{\dag}c_{A,2,\downarrow}\right)-\left(c_{A,2,\downarrow}^{\dag}c_{B,1,\uparrow}-c_{B,2,\downarrow}^{\dag}c_{A,1,\uparrow}\right)\\
\frac{1}{2}\begin{bmatrix}\left(c_{A,1,\uparrow}^{\dag}c_{B,2,\uparrow}-c_{A,1,\downarrow}^{\dag}c_{B,2,\downarrow}-c_{B,1,\uparrow}^{\dag}c_{A,2,\uparrow}+c_{B,1,\downarrow}^{\dag}c_{A,2,\downarrow}\right)\\
-\left(c_{A,2,\uparrow}^{\dagger}c_{B,1,\uparrow}-c_{A,2,\downarrow}^{\dag}c_{B,1,\downarrow}-c_{B,2,\uparrow}^{\dag}c_{A,1,\uparrow}+c_{B,2,\downarrow}^{\dag}c_{A,1,\downarrow}\right)
\end{bmatrix}\\
i\left[\left(c_{A,2,\uparrow}^{\dag}c_{B,2,\downarrow}^{\dag}-c_{A,2,\downarrow}^{\dag}c_{B,2,\uparrow}^{\dag}\right)+\left(c_{A,2,\uparrow}c_{B,2,\downarrow}-c_{A,2,\downarrow}c_{B,2,\uparrow}\right)\right]\\
\left(c_{A,1,\uparrow}^{\dag}c_{B,1,\downarrow}^{\dag}-c_{A,1,\downarrow}^{\dag}c_{B,1,\uparrow}^{\dag}\right)-\left(c_{A,1,\uparrow}c_{B,1,\downarrow}-c_{A,1,\downarrow}c_{B,1,\uparrow}\right)
\end{pmatrix},
    \end{aligned}
\end{equation}

\begin{equation}
    \begin{aligned}
        \tilde{B}_{10}=&B_{18}-B_{17}\\\sim&\begin{pmatrix}i\begin{bmatrix}\left[\left(c_{A,1,\uparrow}^{\dag}c_{A,1,\downarrow}-c_{A,2,\uparrow}^{\dag}c_{A,2,\downarrow}\right)-\left(c_{B,1,\uparrow}^{\dag}c_{B,1,\downarrow}-c_{B,2,\uparrow}^{\dag}c_{B,2,\downarrow}\right)\right]\\
-\left[\left(c_{A,1,\downarrow}^{\dag}c_{A,1,\uparrow}-c_{A,2,\downarrow}^{\dag}c_{A,2,\uparrow}\right)-\left(c_{B,1,\downarrow}^{\dag}c_{B,1,\uparrow}-c_{B,2,\downarrow}^{\dag}c_{B,2,\uparrow}\right)\right]
\end{bmatrix}\\
\begin{bmatrix}\left(c_{A,1,\uparrow}^{\dag}c_{A,1,\downarrow}-c_{A,2,\uparrow}^{\dag}c_{A,2,\downarrow}\right)-\left(c_{B,1,\uparrow}^{\dag}c_{B,1,\downarrow}-c_{B,2,\uparrow}^{\dag}c_{B,2,\downarrow}\right)\\
+\left(c_{A,1,\downarrow}^{\dag}c_{A,1,\uparrow}-c_{A,2,\downarrow}^{\dag}c_{A,2,\uparrow}\right)-\left(c_{B,1,\downarrow}^{\dag}c_{B,1,\uparrow}-c_{B,2,\downarrow}^{\dag}c_{B,2,\uparrow}\right)
\end{bmatrix}\\
i\begin{bmatrix}\left(c_{A,1,\uparrow}^{\dag}c_{A,2,\downarrow}^{\dag}-c_{A,1,\downarrow}^{\dag}c_{A,2,\uparrow}^{\dag}\right)+\left(c_{B,1,\uparrow}^{\dag}c_{B,2,\downarrow}^{\dag}-c_{B,1,\downarrow}^{\dag}c_{B,2,\uparrow}^{\dag}\right)\\
+\left(c_{A,1,\uparrow}c_{A,2,\downarrow}-c_{A,1,\downarrow}c_{A,2,\uparrow}\right)+\left(c_{B,1,\uparrow}c_{B,2,\downarrow}-c_{B,1,\downarrow}c_{B,2,\uparrow}\right)
\end{bmatrix}\\
\begin{bmatrix}\left(c_{A,1,\uparrow}^{\dag}c_{A,2,\downarrow}^{\dag}-c_{A,1,\downarrow}^{\dag}c_{A,2,\uparrow}^{\dag}\right)+\left(c_{B,1,\uparrow}^{\dag}c_{B,2,\downarrow}^{\dag}-c_{B,1,\downarrow}^{\dag}c_{B,2,\uparrow}^{\dag}\right)\\
-\left(c_{A,1,\uparrow}c_{A,2,\downarrow}-c_{A,1,\downarrow}c_{A,2,\uparrow}\right)-\left(c_{B,1,\uparrow}c_{B,2,\downarrow}-c_{B,1,\downarrow}c_{B,2,\uparrow}\right)
\end{bmatrix}\\
\begin{bmatrix}\left(c_{A,1,\uparrow}^{\dag}c_{A,1,\uparrow}-c_{A,1,\downarrow}^{\dag}c_{A,1,\downarrow}-c_{A,2,\uparrow}^{\dag}c_{A,2,\uparrow}+c_{A,2,\downarrow}^{\dag}c_{A,2,\downarrow}\right)\\
-\left(c_{B,1,\uparrow}^{\dag}c_{B,1,\uparrow}-c_{B,1,\downarrow}^{\dag}c_{B,1,\downarrow}-c_{B,2,\uparrow}^{\dag}c_{B,2,\uparrow}+c_{B,2,\downarrow}^{\dag}c_{B,2,\downarrow}\right)
\end{bmatrix}
\end{pmatrix},
    \end{aligned}
\end{equation}

\begin{equation} 
    \begin{aligned}
        \tilde{B}_{11}=&B_{18}+B_{17}\\\sim&\begin{pmatrix}i\begin{bmatrix}\left[\left(c_{A,1,\uparrow}^{\dag}c_{A,1,\downarrow}-c_{A,2,\uparrow}^{\dag}c_{A,2,\downarrow}\right)+\left(c_{B,1,\uparrow}^{\dag}c_{B,1,\downarrow}-c_{B,2,\uparrow}^{\dag}c_{B,2,\downarrow}\right)\right]\\
-\left[\left(c_{A,1,\downarrow}^{\dag}c_{A,1,\uparrow}-c_{A,2,\downarrow}^{\dag}c_{A,2,\uparrow}\right)+\left(c_{B,1,\downarrow}^{\dag}c_{B,1,\uparrow}-c_{B,2,\downarrow}^{\dag}c_{B,2,\uparrow}\right)\right]
\end{bmatrix}\\
\begin{bmatrix}\left(c_{A,1,\uparrow}^{\dag}c_{A,1,\downarrow}-c_{A,2,\uparrow}^{\dag}c_{A,2,\downarrow}\right)+\left(c_{B,1,\uparrow}^{\dag}c_{B,1,\downarrow}-c_{B,2,\uparrow}^{\dag}c_{B,2,\downarrow}\right)\\
+\left(c_{A,1,\downarrow}^{\dag}c_{A,1,\uparrow}-c_{A,2,\downarrow}^{\dag}c_{A,2,\uparrow}\right)+\left(c_{B,1,\downarrow}^{\dag}c_{B,1,\uparrow}-c_{B,2,\downarrow}^{\dag}c_{B,2,\uparrow}\right)
\end{bmatrix}\\
i\begin{bmatrix}\left(c_{A,1,\uparrow}^{\dag}c_{A,2,\downarrow}^{\dag}-c_{A,1,\downarrow}^{\dag}c_{A,2,\uparrow}^{\dag}\right)-\left(c_{B,1,\uparrow}^{\dag}c_{B,2,\downarrow}^{\dag}-c_{B,1,\downarrow}^{\dag}c_{B,2,\uparrow}^{\dag}\right)\\
+\left(c_{A,1,\uparrow}c_{A,2,\downarrow}-c_{A,1,\downarrow}c_{A,2,\uparrow}\right)-\left(c_{B,1,\uparrow}c_{B,2,\downarrow}-c_{B,1,\downarrow}c_{B,2,\uparrow}\right)
\end{bmatrix}\\
\begin{bmatrix}\left[\left(c_{A,1,\uparrow}^{\dag}c_{A,2,\downarrow}^{\dag}-c_{A,1,\downarrow}^{\dag}c_{A,2,\uparrow}^{\dag}\right)-\left(c_{B,1,\uparrow}^{\dag}c_{B,2,\downarrow}^{\dag}-c_{B,1,\downarrow}^{\dag}c_{B,2,\uparrow}^{\dag}\right)\right]\\
-\left[\left(c_{A,1,\uparrow}c_{A,2,\downarrow}-c_{A,1,\downarrow}c_{A,2,\uparrow}\right)-\left(c_{B,1,\uparrow}c_{B,2,\downarrow}-c_{B,1,\downarrow}c_{B,2,\uparrow}\right)\right]
\end{bmatrix}\\
\begin{bmatrix}\left(c_{A,1,\uparrow}^{\dag}c_{A,1,\uparrow}-c_{A,1,\downarrow}^{\dag}c_{A,1,\downarrow}-c_{A,2,\uparrow}^{\dag}c_{A,2,\uparrow}+c_{A,2,\downarrow}^{\dag}c_{A,2,\downarrow}\right)\\
+\left(c_{B,1,\uparrow}^{\dag}c_{B,1,\uparrow}-c_{B,1,\downarrow}^{\dag}c_{B,1,\downarrow}-c_{B,2,\uparrow}^{\dag}c_{B,2,\uparrow}+c_{B,2,\downarrow}^{\dag}c_{B,2,\downarrow}\right)
\end{bmatrix}
\end{pmatrix},
    \end{aligned}
\end{equation}

\begin{equation}
    \begin{aligned}
        \tilde{B}_{12}=&B_{19}-B_{20}\\\sim&\begin{pmatrix}i\begin{bmatrix}\left(c_{A,1,\uparrow}^{\dag}c_{B,1,\downarrow}+c_{B,2,\uparrow}^{\dag}c_{A,2,\downarrow}\right)-\left(c_{A,2,\uparrow}^{\dag}c_{B,2,\downarrow}+c_{B,1,\uparrow}^{\dag}c_{A,1,\downarrow}\right)\\
+\left(c_{A,1,\downarrow}^{\dag}c_{B,1,\uparrow}+c_{B,2,\downarrow}^{\dag}c_{A,2,\uparrow}\right)-\left(c_{A,2,\downarrow}^{\dag}c_{B,2,\uparrow}+c_{B,1,\downarrow}^{\dag}c_{A,1,\uparrow}\right)
\end{bmatrix}\\
\begin{bmatrix}\left[\left(c_{A,1,\uparrow}^{\dag}c_{B,1,\downarrow}+c_{B,2,\uparrow}^{\dag}c_{A,2,\downarrow}\right)-\left(c_{A,2,\uparrow}^{\dag}c_{B,2,\downarrow}+c_{B,1,\uparrow}^{\dag}c_{A,1,\downarrow}\right)\right]\\
-\left[\left(c_{A,1,\downarrow}^{\dag}c_{B,1,\uparrow}+c_{B,2,\downarrow}^{\dag}c_{A,2,\uparrow}\right)-\left(c_{A,2,\downarrow}^{\dag}c_{B,2,\uparrow}+c_{B,1,\downarrow}^{\dag}c_{A,1,\uparrow}\right)\right]
\end{bmatrix}\\
i\begin{bmatrix}\left(c_{A,1,\uparrow}^{\dag}c_{B,2,\downarrow}^{\dag}-c_{A,1,\downarrow}^{\dag}c_{B,2,\uparrow}^{\dag}\right)+\left(c_{A,2,\uparrow}^{\dag}c_{B,1,\downarrow}^{\dag}-c_{A,2,\downarrow}^{\dag}c_{B,1,\uparrow}^{\dag}\right)\\
+\left(c_{A,2,\uparrow}c_{B,1,\downarrow}-c_{A,2,\downarrow}c_{B,1,\uparrow}\right)+\left(c_{A,1,\uparrow}c_{B,2,\downarrow}-c_{A,1,\downarrow}c_{B,2,\uparrow}\right)
\end{bmatrix}\\
\begin{bmatrix}\left(c_{A,1,\uparrow}^{\dag}c_{B,2,\downarrow}^{\dag}-c_{A,1,\downarrow}^{\dag}c_{B,2,\uparrow}^{\dag}\right)+\left(c_{A,2,\uparrow}^{\dag}c_{B,1,\downarrow}^{\dag}-c_{A,2,\downarrow}^{\dag}c_{B,1,\uparrow}^{\dag}\right)\\
-\left(c_{A,2,\uparrow}c_{B,1,\downarrow}-c_{A,2,\downarrow}c_{B,1,\uparrow}\right)-\left(c_{A,1,\uparrow}c_{B,2,\downarrow}-c_{A,1,\downarrow}c_{B,2,\uparrow}\right)
\end{bmatrix}\\
i\begin{bmatrix}\left(c_{A,1,\uparrow}^{\dag}c_{B,1,\uparrow}-c_{A,1,\downarrow}^{\dag}c_{B,1,\downarrow}+c_{B,2,\uparrow}^{\dag}c_{A,2,\uparrow}-c_{B,2,\downarrow}^{\dag}c_{A,2,\downarrow}\right)\\
-\left(c_{A,2,\uparrow}^{\dag}c_{B,2,\uparrow}-c_{A,2,\downarrow}^{\dag}c_{B,2,\downarrow}+c_{B,1,\uparrow}^{\dag}c_{A,1,\uparrow}-c_{B,1,\downarrow}^{\dag}c_{A,1,\downarrow}\right)
\end{bmatrix}
\end{pmatrix},
    \end{aligned}
\end{equation}

\begin{equation}
    \begin{aligned}
        \tilde{B}_{13}=&B_{19}+B_{20}\\\sim&\begin{pmatrix}i\begin{bmatrix}\left[\left(c_{A,1,\uparrow}^{\dag}c_{B,1,\downarrow}+c_{B,2,\uparrow}^{\dag}c_{A,2,\downarrow}\right)+\left(c_{A,2,\uparrow}^{\dag}c_{B,2,\downarrow}+c_{B,1,\uparrow}^{\dag}c_{A,1,\downarrow}\right)\right]\\
-\left[\left(c_{A,1,\downarrow}^{\dag}c_{B,1,\uparrow}+c_{B,2,\downarrow}^{\dag}c_{A,2,\uparrow}\right)+\left(c_{A,2,\downarrow}^{\dag}c_{B,2,\uparrow}+c_{B,1,\downarrow}^{\dag}c_{A,1,\uparrow}\right)\right]
\end{bmatrix}\\
\begin{bmatrix}\left(c_{A,1,\uparrow}^{\dag}c_{B,1,\downarrow}+c_{B,2,\uparrow}^{\dag}c_{A,2,\downarrow}\right)+\left(c_{A,2,\uparrow}^{\dag}c_{B,2,\downarrow}+c_{B,1,\uparrow}^{\dag}c_{A,1,\downarrow}\right)\\
+\left(c_{A,1,\downarrow}^{\dag}c_{B,1,\uparrow}+c_{B,2,\downarrow}^{\dag}c_{A,2,\uparrow}\right)+\left(c_{A,2,\downarrow}^{\dag}c_{B,2,\uparrow}+c_{B,1,\downarrow}^{\dag}c_{A,1,\uparrow}\right)
\end{bmatrix}\\
i\begin{bmatrix}\left(c_{A,1,\uparrow}^{\dag}c_{B,2,\downarrow}^{\dag}-c_{A,1,\downarrow}^{\dag}c_{B,2,\uparrow}^{\dag}\right)-\left(c_{A,2,\uparrow}^{\dag}c_{B,1,\downarrow}^{\dag}-c_{A,2,\downarrow}^{\dag}c_{B,1,\uparrow}^{\dag}\right)\\
-\left(c_{A,2,\uparrow}c_{B,1,\downarrow}-c_{A,2,\downarrow}c_{B,1,\uparrow}\right)+\left(c_{A,1,\uparrow}c_{B,2,\downarrow}-c_{A,1,\downarrow}c_{B,2,\uparrow}\right)
\end{bmatrix}\\
\begin{bmatrix}\left(c_{A,1,\uparrow}^{\dag}c_{B,2,\downarrow}^{\dag}-c_{A,1,\downarrow}^{\dag}c_{B,2,\uparrow}^{\dag}\right)-\left(c_{A,2,\uparrow}^{\dag}c_{B,1,\downarrow}^{\dag}-c_{A,2,\downarrow}^{\dag}c_{B,1,\uparrow}^{\dag}\right)\\
+\left(c_{A,2,\uparrow}c_{B,1,\downarrow}-c_{A,2,\downarrow}c_{B,1,\uparrow}\right)-\left(c_{A,1,\uparrow}c_{B,2,\downarrow}-c_{A,1,\downarrow}c_{B,2,\uparrow}\right)
\end{bmatrix}\\
\begin{bmatrix}\left(c_{A,1,\uparrow}^{\dag}c_{B,1,\uparrow}-c_{A,1,\downarrow}^{\dag}c_{B,1,\downarrow}+c_{B,2,\uparrow}^{\dag}c_{A,2,\uparrow}-c_{B,2,\downarrow}^{\dag}c_{A,2,\downarrow}\right)\\
+\left(c_{A,2,\uparrow}^{\dag}c_{B,2,\uparrow}-c_{A,2,\downarrow}^{\dag}c_{B,2,\downarrow}+c_{B,1,\uparrow}^{\dag}c_{A,1,\uparrow}-c_{B,1,\downarrow}^{\dag}c_{A,1,\downarrow}\right)
\end{bmatrix}
\end{pmatrix},
    \end{aligned}
\end{equation}

\begin{equation}
    \begin{aligned}
        \tilde{B}_{14}=&B_{21}\oplus B_{22}\\\sim&\begin{pmatrix}\frac{i}{2}\left[\left(c_{A,1,\uparrow}^{\dag}c_{B,2,\uparrow}+c_{B,1,\uparrow}^{\dag}c_{A,2,\uparrow}\right)-\left(c_{A,2,\uparrow}^{\dag}c_{B,1,\uparrow}+c_{B,2,\uparrow}^{\dag}c_{A,1,\uparrow}\right)\right]\\
\frac{i}{2}\left[\left(c_{A,1,\downarrow}^{\dag}c_{B,2,\downarrow}+c_{B,1,\downarrow}^{\dag}c_{A,2,\downarrow}\right)-\left(c_{A,2,\downarrow}^{\dag}c_{B,1,\downarrow}+c_{B,2,\downarrow}^{\dag}c_{A,1,\downarrow}\right)\right]\\
\frac{i}{2}\left[\left(c_{A,1,\uparrow}^{\dag}c_{B,2,\downarrow}+c_{B,1,\uparrow}^{\dag}c_{A,2,\downarrow}\right)-\left(c_{A,2,\downarrow}^{\dag}c_{B,1,\uparrow}+c_{B,2,\downarrow}^{\dag}c_{A,1,\uparrow}\right)\right]\\
\frac{i}{2}\left[\left(c_{A,1,\downarrow}^{\dag}c_{B,2,\uparrow}+c_{B,1,\downarrow}^{\dag}c_{A,2,\uparrow}\right)-\left(c_{A,2,\uparrow}^{\dag}c_{B,1,\downarrow}+c_{B,2,\uparrow}^{\dag}c_{A,1,\downarrow}\right)\right]\\
i\left(c_{A,1,\uparrow}^{\dag}c_{B,1,\uparrow}^{\dag}+c_{A,1,\uparrow}c_{B,1,\uparrow}\right)\\
i\left(c_{A,1,\downarrow}^{\dag}c_{B,1,\downarrow}^{\dag}+c_{A,1,\downarrow}c_{B,1,\downarrow}\right)\\
c_{A,2,\uparrow}^{\dag}c_{B,2,\uparrow}^{\dag}-c_{A,2,\uparrow}c_{B,2,\uparrow}\\
c_{A,2,\downarrow}^{\dag}c_{B,2,\downarrow}^{\dag}-c_{A,2,\downarrow}c_{B,2,\downarrow}\\
\frac{i}{2}\left[\left(c_{A,1,\uparrow}^{\dag}c_{B,1,\downarrow}^{\dag}+c_{A,1,\downarrow}^{\dag}c_{B,1,\uparrow}^{\dag}\right)+\left(c_{A,1,\uparrow}c_{B,1,\downarrow}+c_{A,1,\downarrow}c_{B,1,\uparrow}\right)\right]\\
\frac{1}{2}\left[\left(c_{A,2,\uparrow}^{\dag}c_{B,2,\downarrow}^{\dag}+c_{A,2,\downarrow}^{\dag}c_{B,2,\uparrow}^{\dag}\right)-\left(c_{A,2,\uparrow}c_{B,2,\downarrow}+c_{A,2,\downarrow}c_{B,2,\uparrow}\right)\right]\\
\frac{1}{2}\left[\left(c_{A,1,\uparrow}^{\dag}c_{B,2,\uparrow}+c_{B,1,\uparrow}^{\dag}c_{A,2,\uparrow}\right)+\left(c_{A,2,\uparrow}^{\dag}c_{B,1,\uparrow}+c_{B,2,\uparrow}^{\dag}c_{A,1,\uparrow}\right)\right]\\
\frac{1}{2}\left[\left(c_{A,1,\downarrow}^{\dag}c_{B,2,\downarrow}+c_{B,1,\downarrow}^{\dag}c_{A,2,\downarrow}\right)+\left(c_{A,2,\downarrow}^{\dag}c_{B,1,\downarrow}+c_{B,2,\downarrow}^{\dag}c_{A,1,\downarrow}\right)\right]\\
\frac{1}{2}\left[\left(c_{A,1,\downarrow}^{\dag}c_{B,2,\uparrow}+c_{B,1,\downarrow}^{\dag}c_{A,2,\uparrow}\right)+\left(c_{A,2,\uparrow}^{\dag}c_{B,1,\downarrow}+c_{B,2,\uparrow}^{\dag}c_{A,1,\downarrow}\right)\right]\\
\frac{1}{2}\left[\left(c_{A,1,\uparrow}^{\dag}c_{B,2,\downarrow}+c_{B,1,\uparrow}^{\dag}c_{A,2,\downarrow}\right)+\left(c_{A,2,\downarrow}^{\dag}c_{B,1,\uparrow}+c_{B,2,\downarrow}^{\dag}c_{A,1,\uparrow}\right)\right]\\
i\left(c_{A,2,\uparrow}^{\dag}c_{B,2,\uparrow}^{\dag}+c_{A,2,\uparrow}c_{B,2,\uparrow}\right)\\
i\left(c_{A,2,\downarrow}^{\dag}c_{B,2,\downarrow}^{\dag}+c_{A,2,\downarrow}c_{B,2,\downarrow}\right)\\
c_{A,1,\uparrow}^{\dag}c_{B,1,\uparrow}^{\dag}-c_{A,1,\uparrow}c_{B,1,\uparrow}\\
c_{A,1,\downarrow}^{\dag}c_{B,1,\downarrow}^{\dag}-c_{A,1,\downarrow}c_{B,1,\downarrow}\\
\frac{i}{2}\left[\left(c_{A,2,\uparrow}^{\dag}c_{B,2,\downarrow}^{\dag}+c_{A,2,\downarrow}^{\dag}c_{B,2,\uparrow}^{\dag}\right)+\left(c_{A,2,\uparrow}c_{B,2,\downarrow}+c_{A,2,\downarrow}c_{B,2,\uparrow}\right)\right]\\
\frac{1}{2}\left[\left(c_{A,1,\uparrow}^{\dag}c_{B,1,\downarrow}^{\dag}+c_{A,1,\downarrow}^{\dag}c_{B,1,\uparrow}^{\dag}\right)-\left(c_{A,1,\uparrow}c_{B,1,\downarrow}+c_{A,1,\downarrow}c_{B,1,\uparrow}\right)\right]
\end{pmatrix},
    \end{aligned}
\end{equation}

\begin{equation}
    \begin{aligned}
        \tilde{B}_{15}=&B_{24}-B_{23}\\\sim&\begin{pmatrix}\frac{i}{2}\begin{bmatrix}\left[\left(c_{A,1,\uparrow}^{\dag}c_{A,1,\downarrow}+c_{A,2,\uparrow}^{\dag}c_{A,2,\downarrow}\right)-\left(c_{B,1,\uparrow}^{\dag}c_{B,1,\downarrow}+c_{B,2,\uparrow}^{\dag}c_{B,2,\downarrow}\right)\right]\\
-\left[\left(c_{A,1,\downarrow}^{\dag}c_{A,1,\uparrow}+c_{A,2,\downarrow}^{\dag}c_{A,2,\uparrow}\right)-\left(c_{B,1,\downarrow}^{\dag}c_{B,1,\uparrow}+c_{B,2,\downarrow}^{\dag}c_{B,2,\uparrow}\right)\right]
\end{bmatrix}\\
\frac{1}{2}\begin{bmatrix}\left(c_{A,1,\uparrow}^{\dag}c_{A,1,\downarrow}+c_{A,2,\uparrow}^{\dag}c_{A,2,\downarrow}\right)-\left(c_{B,1,\uparrow}^{\dag}c_{B,1,\downarrow}+c_{B,2,\uparrow}^{\dag}c_{B,2,\downarrow}\right)\\
+\left(c_{A,1,\downarrow}^{\dag}c_{A,1,\uparrow}+c_{A,2,\downarrow}^{\dag}c_{A,2,\uparrow}\right)-\left(c_{B,1,\downarrow}^{\dag}c_{B,1,\uparrow}+c_{B,2,\downarrow}^{\dag}c_{B,2,\uparrow}\right)
\end{bmatrix}\\
i\left[\left(c_{A,1,\uparrow}^{\dag}c_{A,2,\uparrow}^{\dag}+c_{B,1,\uparrow}^{\dag}c_{B,2,\uparrow}^{\dag}\right)+\left(c_{A,1,\uparrow}c_{A,2,\uparrow}+c_{B,1,\uparrow}c_{B,2,\uparrow}\right)\right]\\
i\left[\left(c_{A,1,\downarrow}^{\dag}c_{A,2,\downarrow}^{\dag}+c_{B,1,\downarrow}^{\dag}c_{B,2,\downarrow}^{\dag}\right)+\left(c_{A,1,\downarrow}c_{A,2,\downarrow}+c_{B,1,\downarrow}c_{B,2,\downarrow}\right)\right]\\
\left(c_{A,1,\uparrow}^{\dag}c_{A,2,\uparrow}^{\dag}+c_{B,1,\uparrow}^{\dag}c_{B,2,\uparrow}^{\dag}\right)-\left(c_{A,1,\uparrow}c_{A,2,\uparrow}+c_{B,1,\uparrow}c_{B,2,\uparrow}\right)\\
\left(c_{A,1,\downarrow}^{\dag}c_{A,2,\downarrow}^{\dag}+c_{B,1,\downarrow}^{\dag}c_{B,2,\downarrow}^{\dag}\right)-\left(c_{A,1,\downarrow}c_{A,2,\downarrow}+c_{B,1,\downarrow}c_{B,2,\downarrow}\right)\\
\frac{i}{2}\begin{bmatrix}\left(c_{A,1,\uparrow}^{\dag}c_{A,2,\downarrow}^{\dag}+c_{A,1,\downarrow}^{\dag}c_{A,2,\uparrow}^{\dag}\right)+\left(c_{B,1,\uparrow}^{\dag}c_{B,2,\downarrow}^{\dag}+c_{B,1,\downarrow}^{\dag}c_{B,2,\uparrow}^{\dag}\right)\\
+\left(c_{A,1,\uparrow}c_{A,2,\downarrow}+c_{A,1,\downarrow}c_{A,2,\uparrow}\right)+\left(c_{B,1,\uparrow}c_{B,2,\downarrow}+c_{B,1,\downarrow}c_{B,2,\uparrow}\right)
\end{bmatrix}\\
\frac{1}{2}\begin{bmatrix}\left[\left(c_{A,1,\uparrow}^{\dag}c_{A,2,\downarrow}^{\dag}+c_{A,1,\downarrow}^{\dag}c_{A,2,\uparrow}^{\dag}\right)+\left(c_{B,1,\uparrow}^{\dag}c_{B,2,\downarrow}^{\dag}+c_{B,1,\downarrow}^{\dag}c_{B,2,\uparrow}^{\dag}\right)\right]\\
-\left(c_{A,1,\uparrow}c_{A,2,\downarrow}+c_{A,1,\downarrow}c_{A,2,\uparrow}\right)-\left(c_{B,1,\uparrow}c_{B,2,\downarrow}+c_{B,1,\downarrow}c_{B,2,\uparrow}\right)
\end{bmatrix}\\
\left(c_{A,1,\uparrow}^{\dag}c_{A,1,\uparrow}+c_{A,2,\uparrow}^{\dag}c_{A,2,\uparrow}\right)-\left(c_{B,1,\uparrow}^{\dag}c_{B,1,\uparrow}+c_{B,2,\uparrow}^{\dag}c_{B,2,\uparrow}\right)\\
\left(c_{A,1,\downarrow}^{\dag}c_{A,1,\downarrow}+c_{A,2,\downarrow}^{\dag}c_{A,2,\downarrow}\right)-\left(c_{B,1,\downarrow}^{\dag}c_{B,1,\downarrow}+c_{B,2,\downarrow}^{\dag}c_{B,2,\downarrow}\right)
\end{pmatrix},
    \end{aligned}
\end{equation}

\begin{equation}
    \begin{aligned}
        \tilde{B}_{16}=&B_{24}+B_{23}\\\sim&\begin{pmatrix}\frac{i}{2}\begin{bmatrix}\left[\left(c_{A,1,\uparrow}^{\dag}c_{A,1,\downarrow}+c_{A,2,\uparrow}^{\dag}c_{A,2,\downarrow}\right)+\left(c_{B,1,\uparrow}^{\dag}c_{B,1,\downarrow}+c_{B,2,\uparrow}^{\dag}c_{B,2,\downarrow}\right)\right]\\
-\left[\left(c_{A,1,\downarrow}^{\dag}c_{A,1,\uparrow}+c_{A,2,\downarrow}^{\dag}c_{A,2,\uparrow}\right)+\left(c_{B,1,\downarrow}^{\dag}c_{B,1,\uparrow}+c_{B,2,\downarrow}^{\dag}c_{B,2,\uparrow}\right)\right]
\end{bmatrix}\\
\frac{1}{2}\begin{bmatrix}\left(c_{A,1,\uparrow}^{\dag}c_{A,1,\downarrow}+c_{A,2,\uparrow}^{\dag}c_{A,2,\downarrow}\right)+\left(c_{B,1,\uparrow}^{\dag}c_{B,1,\downarrow}+c_{B,2,\uparrow}^{\dag}c_{B,2,\downarrow}\right)\\
+\left(c_{A,1,\downarrow}^{\dag}c_{A,1,\uparrow}+c_{A,2,\downarrow}^{\dag}c_{A,2,\uparrow}\right)+\left(c_{B,1,\downarrow}^{\dag}c_{B,1,\uparrow}+c_{B,2,\downarrow}^{\dag}c_{B,2,\uparrow}\right)
\end{bmatrix}\\
i\left[\left(c_{A,1,\uparrow}^{\dag}c_{A,2,\uparrow}^{\dag}-c_{B,1,\uparrow}^{\dag}c_{B,2,\uparrow}^{\dag}\right)+\left(c_{A,1,\uparrow}c_{A,2,\uparrow}-c_{B,1,\uparrow}c_{B,2,\uparrow}\right)\right]\\
i\left[\left(c_{A,1,\downarrow}^{\dag}c_{A,2,\downarrow}^{\dag}-c_{B,1,\downarrow}^{\dag}c_{B,2,\downarrow}^{\dag}\right)+\left(c_{A,1,\downarrow}c_{A,2,\downarrow}-c_{B,1,\downarrow}c_{B,2,\downarrow}\right)\right]\\
\left(c_{A,1,\uparrow}^{\dag}c_{A,2,\uparrow}^{\dag}-c_{B,1,\uparrow}^{\dag}c_{B,2,\uparrow}^{\dag}\right)-\left(c_{A,1,\uparrow}c_{A,2,\uparrow}-c_{B,1,\uparrow}c_{B,2,\uparrow}\right)\\
\left(c_{A,1,\downarrow}^{\dag}c_{A,2,\downarrow}^{\dag}-c_{B,1,\downarrow}^{\dag}c_{B,2,\downarrow}^{\dag}\right)-\left(c_{A,1,\downarrow}c_{A,2,\downarrow}-c_{B,1,\downarrow}c_{B,2,\downarrow}\right)\\
\frac{i}{2}\begin{bmatrix}\left(c_{A,1,\uparrow}^{\dag}c_{A,2,\downarrow}^{\dag}+c_{A,1,\downarrow}^{\dag}c_{A,2,\uparrow}^{\dag}\right)-\left(c_{B,1,\uparrow}^{\dag}c_{B,2,\downarrow}^{\dag}+c_{B,1,\downarrow}^{\dag}c_{B,2,\uparrow}^{\dag}\right)\\
+\left(c_{A,1,\uparrow}c_{A,2,\downarrow}+c_{A,1,\downarrow}c_{A,2,\uparrow}\right)-\left(c_{B,1,\uparrow}c_{B,2,\downarrow}+c_{B,1,\downarrow}c_{B,2,\uparrow}\right)
\end{bmatrix}\\
\frac{1}{2}\begin{bmatrix}\left[\left(c_{A,1,\uparrow}^{\dag}c_{A,2,\downarrow}^{\dag}+c_{A,1,\downarrow}^{\dag}c_{A,2,\uparrow}^{\dag}\right)-\left(c_{B,1,\uparrow}^{\dag}c_{B,2,\downarrow}^{\dag}+c_{B,1,\downarrow}^{\dag}c_{B,2,\uparrow}^{\dag}\right)\right]\\
-\left(c_{A,1,\uparrow}c_{A,2,\downarrow}+c_{A,1,\downarrow}c_{A,2,\uparrow}\right)+\left(c_{B,1,\uparrow}c_{B,2,\downarrow}+c_{B,1,\downarrow}c_{B,2,\uparrow}\right)
\end{bmatrix}\\
\left(c_{A,1,\uparrow}^{\dag}c_{A,1,\uparrow}+c_{A,2,\uparrow}^{\dag}c_{A,2,\uparrow}-1\right)+\left(c_{B,1,\uparrow}^{\dag}c_{B,1,\uparrow}+c_{B,2,\uparrow}^{\dag}c_{B,2,\uparrow}-1\right)\\
\left(c_{A,1,\downarrow}^{\dag}c_{A,1,\downarrow}+c_{A,2,\downarrow}^{\dag}c_{A,2,\downarrow}-1\right)+\left(c_{B,1,\downarrow}^{\dag}c_{B,1,\downarrow}+c_{B,2,\downarrow}^{\dag}c_{B,2,\downarrow}-1\right)
\end{pmatrix},
    \end{aligned}
\end{equation}

\begin{equation}
    \begin{aligned}
        \tilde{B}_{17}=&B_{26}-B_{25}\\\sim&\begin{pmatrix}\frac{1}{2}\begin{bmatrix}\left[\left(c_{A,1,\uparrow}^{\dag}c_{B,1,\downarrow}-c_{B,2,\uparrow}^{\dag}c_{A,2,\downarrow}\right)+\left(c_{A,2,\uparrow}^{\dag}c_{B,2,\downarrow}-c_{B,1,\uparrow}^{\dag}c_{A,1,\downarrow}\right)\right]\\
-\left(c_{A,1,\downarrow}^{\dag}c_{B,1,\uparrow}-c_{B,2,\downarrow}^{\dag}c_{A,2,\uparrow}\right)-\left(c_{A,2,\downarrow}^{\dag}c_{B,2,\uparrow}-c_{B,1,\downarrow}^{\dag}c_{A,1,\uparrow}\right)
\end{bmatrix}\\
\frac{i}{2}\begin{bmatrix}\left(c_{A,1,\uparrow}^{\dag}c_{B,1,\downarrow}-c_{B,2,\uparrow}^{\dag}c_{A,2,\downarrow}\right)+\left(c_{A,2,\uparrow}^{\dag}c_{B,2,\downarrow}-c_{B,1,\uparrow}^{\dag}c_{A,1,\downarrow}\right)\\
+\left(c_{A,1,\downarrow}^{\dag}c_{B,1,\uparrow}-c_{B,2,\downarrow}^{\dag}c_{A,2,\uparrow}\right)+\left(c_{A,2,\downarrow}^{\dag}c_{B,2,\uparrow}-c_{B,1,\downarrow}^{\dag}c_{A,1,\uparrow}\right)
\end{bmatrix}\\
i\left[\left(c_{A,1,\uparrow}^{\dag}c_{B,2,\uparrow}^{\dag}-c_{A,2,\uparrow}^{\dag}c_{B,1,\uparrow}^{\dag}\right)-\left(c_{A,2,\uparrow}c_{B,1,\uparrow}-c_{A,1,\uparrow}c_{B,2,\uparrow}\right)\right]\\
i\left[\left(c_{A,1,\downarrow}^{\dag}c_{B,2,\downarrow}^{\dag}-c_{A,2,\downarrow}^{\dag}c_{B,1,\downarrow}^{\dag}\right)-\left(c_{A,2,\downarrow}c_{B,1,\downarrow}-c_{A,1,\downarrow}c_{B,2,\downarrow}\right)\right]\\
\left(c_{A,1,\uparrow}^{\dag}c_{B,2,\uparrow}^{\dag}-c_{A,2,\uparrow}^{\dag}c_{B,1,\uparrow}^{\dag}\right)+\left(c_{A,2,\uparrow}c_{B,1,\uparrow}-c_{A,1,\uparrow}c_{B,2,\uparrow}\right)\\
\left(c_{A,1,\downarrow}^{\dag}c_{B,2,\downarrow}^{\dag}-c_{A,2,\downarrow}^{\dag}c_{B,1,\downarrow}^{\dag}\right)+\left(c_{A,2,\downarrow}c_{B,1,\downarrow}-c_{A,1,\downarrow}c_{B,2,\downarrow}\right)\\
\frac{i}{2}\begin{bmatrix}\left[\left(c_{A,1,\uparrow}^{\dag}c_{B,2,\downarrow}^{\dag}+c_{A,1,\downarrow}^{\dag}c_{B,2,\uparrow}^{\dag}\right)-\left(c_{A,2,\uparrow}^{\dag}c_{B,1,\downarrow}^{\dag}+c_{A,2,\downarrow}^{\dag}c_{B,1,\uparrow}^{\dag}\right)\right]\\
-\left[\left(c_{A,2,\uparrow}c_{B,1,\downarrow}+c_{A,2,\downarrow}c_{B,1,\uparrow}\right)-\left(c_{A,1,\uparrow}c_{B,2,\downarrow}+c_{A,1,\downarrow}c_{B,2,\uparrow}\right)\right]
\end{bmatrix}\\
\frac{1}{2}\begin{bmatrix}\left(c_{A,1,\uparrow}^{\dag}c_{B,2,\downarrow}^{\dag}+c_{A,1,\downarrow}^{\dag}c_{B,2,\uparrow}^{\dag}\right)-\left(c_{A,2,\uparrow}^{\dag}c_{B,1,\downarrow}^{\dag}+c_{A,2,\downarrow}^{\dag}c_{B,1,\uparrow}^{\dag}\right)\\
+\left(c_{A,2,\uparrow}c_{B,1,\downarrow}+c_{A,2,\downarrow}c_{B,1,\uparrow}\right)-\left(c_{A,1,\uparrow}c_{B,2,\downarrow}+c_{A,1,\downarrow}c_{B,2,\uparrow}\right)
\end{bmatrix}\\
i\left[\left(c_{A,1,\uparrow}^{\dag}c_{B,1,\uparrow}-c_{B,2,\uparrow}^{\dag}c_{A,2,\uparrow}\right)+\left(c_{A,2,\uparrow}^{\dag}c_{B,2,\uparrow}-c_{B,1,\uparrow}^{\dag}c_{A,1,\uparrow}\right)\right]\\
i\left[\left(c_{A,1,\downarrow}^{\dag}c_{B,1,\downarrow}-c_{B,2,\downarrow}^{\dag}c_{A,2,\downarrow}\right)+\left(c_{A,2,\downarrow}^{\dag}c_{B,2,\downarrow}-c_{B,1,\downarrow}^{\dag}c_{A,1,\downarrow}\right)\right]
\end{pmatrix},
    \end{aligned}
\end{equation}

\begin{equation}
    \begin{aligned}
        \tilde{B}_{18}=&B_{26}+B_{25}\\\sim&\begin{pmatrix}\frac{1}{2}\begin{bmatrix}\left(c_{A,1,\uparrow}^{\dag}c_{B,1,\downarrow}-c_{B,2,\uparrow}^{\dag}c_{A,2,\downarrow}\right)-\left(c_{A,2,\uparrow}^{\dag}c_{B,2,\downarrow}-c_{B,1,\uparrow}^{\dag}c_{A,1,\downarrow}\right)\\
+\left(c_{A,1,\downarrow}^{\dag}c_{B,1,\uparrow}-c_{B,2,\downarrow}^{\dag}c_{A,2,\uparrow}\right)-\left(c_{A,2,\downarrow}^{\dag}c_{B,2,\uparrow}-c_{B,1,\downarrow}^{\dag}c_{A,1,\uparrow}\right)
\end{bmatrix}\\
\frac{i}{2}\begin{bmatrix}\left[\left(c_{A,1,\uparrow}^{\dag}c_{B,1,\downarrow}-c_{B,2,\uparrow}^{\dag}c_{A,2,\downarrow}\right)-\left(c_{A,2,\uparrow}^{\dag}c_{B,2,\downarrow}-c_{B,1,\uparrow}^{\dag}c_{A,1,\downarrow}\right)\right]\\
-\left(c_{A,1,\downarrow}^{\dag}c_{B,1,\uparrow}-c_{B,2,\downarrow}^{\dag}c_{A,2,\uparrow}\right)+\left(c_{A,2,\downarrow}^{\dag}c_{B,2,\uparrow}-c_{B,1,\downarrow}^{\dag}c_{A,1,\uparrow}\right)
\end{bmatrix}\\
i\left[\left(c_{A,1,\uparrow}^{\dag}c_{B,2,\uparrow}^{\dag}+c_{A,2,\uparrow}^{\dag}c_{B,1,\uparrow}^{\dag}\right)+\left(c_{A,2,\uparrow}c_{B,1,\uparrow}+c_{A,1,\uparrow}c_{B,2,\uparrow}\right)\right]\\
i\left[\left(c_{A,1,\downarrow}^{\dag}c_{B,2,\downarrow}^{\dag}+c_{A,2,\downarrow}^{\dag}c_{B,1,\downarrow}^{\dag}\right)+\left(c_{A,2,\downarrow}c_{B,1,\downarrow}+c_{A,1,\downarrow}c_{B,2,\downarrow}\right)\right]\\
\left(c_{A,1,\uparrow}^{\dag}c_{B,2,\uparrow}^{\dag}+c_{A,2,\uparrow}^{\dag}c_{B,1,\uparrow}^{\dag}\right)-\left(c_{A,2,\uparrow}c_{B,1,\uparrow}+c_{A,1,\uparrow}c_{B,2,\uparrow}\right)\\
\left(c_{A,1,\downarrow}^{\dag}c_{B,2,\downarrow}^{\dag}+c_{A,2,\downarrow}^{\dag}c_{B,1,\downarrow}^{\dag}\right)-\left(c_{A,2,\downarrow}c_{B,1,\downarrow}+c_{A,1,\downarrow}c_{B,2,\downarrow}\right)\\
\frac{i}{2}\begin{bmatrix}\left(c_{A,1,\uparrow}^{\dag}c_{B,2,\downarrow}^{\dag}+c_{A,1,\downarrow}^{\dag}c_{B,2,\uparrow}^{\dag}\right)+\left(c_{A,2,\uparrow}^{\dag}c_{B,1,\downarrow}^{\dag}+c_{A,2,\downarrow}^{\dag}c_{B,1,\uparrow}^{\dag}\right)\\
+\left(c_{A,2,\uparrow}c_{B,1,\downarrow}+c_{A,2,\downarrow}c_{B,1,\uparrow}\right)+\left(c_{A,1,\uparrow}c_{B,2,\downarrow}+c_{A,1,\downarrow}c_{B,2,\uparrow}\right)
\end{bmatrix}\\
\frac{1}{2}\begin{bmatrix}\left[\left(c_{A,1,\uparrow}^{\dag}c_{B,2,\downarrow}^{\dag}+c_{A,1,\downarrow}^{\dag}c_{B,2,\uparrow}^{\dag}\right)+\left(c_{A,2,\uparrow}^{\dag}c_{B,1,\downarrow}^{\dag}+c_{A,2,\downarrow}^{\dag}c_{B,1,\uparrow}^{\dag}\right)\right]\\
-\left[\left(c_{A,2,\uparrow}c_{B,1,\downarrow}+c_{A,2,\downarrow}c_{B,1,\uparrow}\right)+\left(c_{A,1,\uparrow}c_{B,2,\downarrow}+c_{A,1,\downarrow}c_{B,2,\uparrow}\right)\right]
\end{bmatrix}\\
\left(c_{A,1,\uparrow}^{\dag}c_{B,1,\uparrow}-c_{B,2,\uparrow}^{\dag}c_{A,2,\uparrow}\right)-\left(c_{A,2,\uparrow}^{\dag}c_{B,2,\uparrow}-c_{B,1,\uparrow}^{\dag}c_{A,1,\uparrow}\right)\\
\left(c_{A,1,\downarrow}^{\dag}c_{B,1,\downarrow}-c_{B,2,\downarrow}^{\dag}c_{A,2,\downarrow}\right)-\left(c_{A,2,\downarrow}^{\dag}c_{B,2,\downarrow}-c_{B,1,\downarrow}^{\dag}c_{A,1,\downarrow}\right)
\end{pmatrix}.
    \end{aligned}
\end{equation}

\end{document}